%% file: main.tex
\documentclass{article}

\usepackage{microtype}
\usepackage{graphicx}
\usepackage{subfigure}

\usepackage{booktabs} % for professional tables

\usepackage{hyperref}

\usepackage[accepted]{icml2024}
\usepackage{amsmath}
\usepackage{amssymb}
\usepackage{amsthm}
\usepackage{xcolor}

\usepackage[capitalize,noabbrev]{cleveref}

\theoremstyle{plain}

\theoremstyle{definition}

\theoremstyle{remark}

\usepackage{times}
\usepackage{latexsym}
\usepackage{blindtext}
\usepackage{microtype}
\usepackage{url}
\usepackage{booktabs}
\usepackage{amsmath}
\usepackage{amssymb} % for checkbox symbol
\usepackage{enumitem} % for customizing item labels
\usepackage{algpseudocode}
\usepackage{graphicx}
\usepackage{multirow}
\usepackage{amssymb}
\usepackage{pifont}
\usepackage{listings}   % For rendering prompt/code templates
\usepackage{xcolor}
\usepackage{float}

\definecolor{light-gray}{gray}{0.95}

\newcommand{\lstbg}[3][0pt]{{\fboxsep#1\colorbox{#2}{\strut #3}}}

\lstdefinelanguage{diff}{
  basicstyle=\ttfamily\small,
  morecomment=[f][\lstbg{red!20}]-,
  morecomment=[f][\lstbg{green!20}]+,
  morecomment=[f][\textit]{@@},
  columns=flexible,
  breaklines=true,
}
\definecolor{light-gray}{gray}{0.95}

\definecolor{darkblue}{rgb}{0, 0, 0.5}
\hypersetup{colorlinks=true, citecolor=darkblue, linkcolor=darkblue, urlcolor=darkblue}

\input{commands}

\makeatletter
\renewenvironment{table}[1][htbp]
  {\@dblfloat{table}[#1]\centering}
  {\end@dblfloat}
\makeatother

\icmltitlerunning{Class-Level Code Generation from Natural Language Using Iterative, Tool-Enhanced Reasoning over Repository}

\begin{document}

\twocolumn[
\icmltitle{Class-Level Code Generation from Natural Language Using Iterative, Tool-Enhanced Reasoning over Repository}



\icmlsetsymbol{equal}{*}

\begin{icmlauthorlist}
\icmlauthor{Ajinkya Deshpande}{equal,yyy}
\icmlauthor{Anmol Agarwal}{equal,yyy}
\icmlauthor{Shashank Shet}{equal,yyy}\\
\icmlauthor{Arun Iyer}{yyy}
\icmlauthor{Aditya Kanade}{yyy}
\icmlauthor{Ramakrishna Bairi}{yyy}
\icmlauthor{Suresh Parthasarathy}{yyy}
\end{icmlauthorlist}

\icmlaffiliation{yyy}{Microsoft Research, India}

\icmlcorrespondingauthor{Ajinkya Deshpande}{ajinkya.deshpande56@gmail.com}
\icmlcorrespondingauthor{Anmol Agarwal}{anmolagarwal4453@gmail.com}
\icmlcorrespondingauthor{Shashank Shet}{shashankshet@gmail.com}
\icmlcorrespondingauthor{Arun Iyer}{ariy@microsoft.com}
\icmlcorrespondingauthor{Aditya Kanade}{kanadeaditya@microsoft.com}
\icmlcorrespondingauthor{Ramakrishna Bairi}{ram.bairi@microsoft.com}
\icmlcorrespondingauthor{Suresh Parthasarathy}{supartha@microsoft.com}

% {ariy,kanadeaditya,ram.bairi,supartha\}@microsoft.com

\icmlkeywords{Machine Learning, ICML}

\vskip 0.3in
]



\printAffiliationsAndNotice{\icmlEqualContribution - Work done while the authors were \href{https://www.microsoft.com/en-us/research/academic-program/research-fellows-program-at-microsoft-research-india/}{Research Fellows} at Microsoft Research India.} % otherwise use the standard text.

\begin{abstract}
% \colorbox{pink}{highlight}\\
% \colorbox{green}{highlight}\\
% \colorbox{yellow}{highlight}\\

Large Language Models (LLMs) have demonstrated significant potential in code generation tasks, achieving promising results at the function or statement level across various benchmarks. However, the complexities associated with creating code artifacts like classes, particularly within the context of real-world software repositories, remain underexplored. Existing research often treats class-level generation as an isolated task, neglecting the intricate dependencies and interactions that characterize real-world software development environments. To address this gap, we introduce \textbf{RepoClassBench}, a comprehensive benchmark designed to rigorously evaluate LLMs in generating complex, class-level code within real-world repositories. RepoClassBench includes "Natural Language to Class generation" tasks across Java, Python and C\#, from a selection of public repositories. We ensure that each class in our dataset not only has cross-file dependencies within the repository but also includes corresponding test cases to verify its functionality. We find that current models struggle with the realistic challenges posed by our benchmark, primarily due to their limited exposure to relevant repository contexts. To address this shortcoming, we introduce \textbf{Retrieve-Repotools-Reflect (RRR)}, a novel approach that equips LLMs with static analysis tools to iteratively navigate and reason about repository-level context in an agent-based framework. Our experiments demonstrate that RRR significantly outperforms existing baselines on RepoClassBench, showcasing its effectiveness across programming languages and under various settings. Our findings emphasize the critical need for code-generation benchmarks to incorporate repository-level dependencies to more accurately reflect the complexities of software development. Furthermore, our work illustrates the benefits of leveraging specialized tools to enhance LLMs' understanding of repository context. We plan to make our dataset and evaluation harness public.

\end{abstract}

\input{sections/introduction}

\begin{figure*}[t]
    \centering
    \scalebox{1}{
    \includegraphics[width=1.0\linewidth]{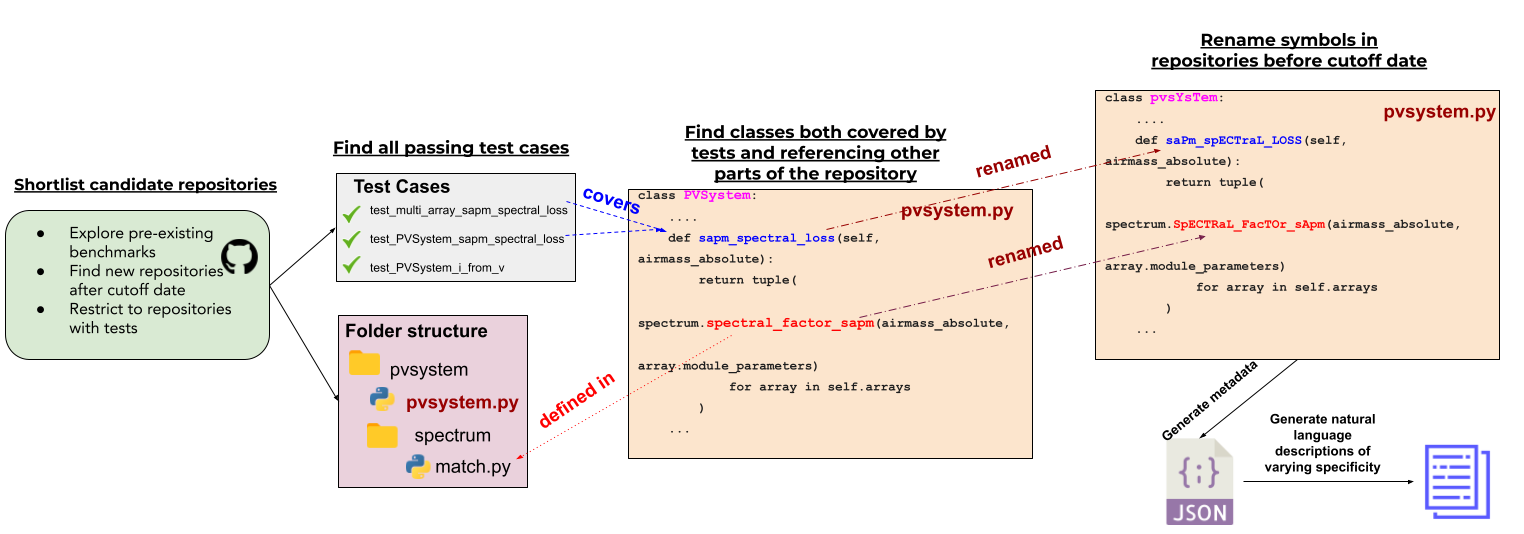}}
    \caption{The dataset creation pipeline involved shortlisting candidate repositories, noting passing test cases, finding classes covered by passing test cases (which make external references) and finally mitigating memorization issues if necessary, using paraphrasing.}
    \label{fig:datasetfigure}
\end{figure*}

\input{sections/related_works}
\input{sections/data_generation_methodology}

\begin{figure*}[t]
    \centering
    \scalebox{1}{
    \includegraphics[width=1.0\linewidth]{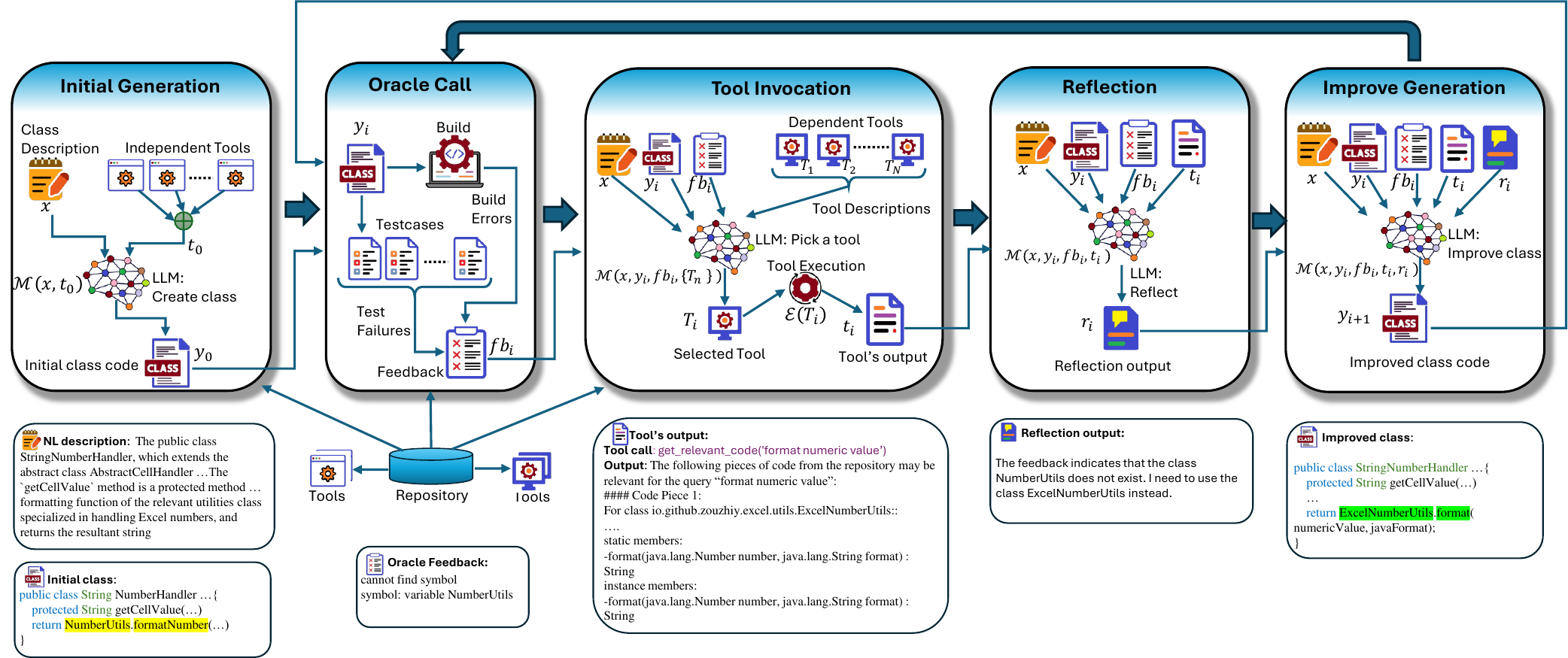}}
    \caption{Flowchart illustrating the procedural framework of \OurMethod. \OurMethod~utilizes the natural language description of the class and outputs of independent tools to create an initial attempt. This attempt is evaluated by an oracle, pinpointing specific errors. Subsequently, \OurMethod~uses repository tools to gather information to rectify the errors. It then reflects on feedback and tool insights to refine the attempt. This iterative cycle persists until either all test cases pass or the maximum allowed number of oracle calls is reached.}
    \label{fig:methodfigure}
\end{figure*}

\input{sections/method}

\input{sections/experimental_results}

\input{sections/conclusion}
\bibliography{references/tool_augmentations,references/context_retrieval,references/benchmarks,references/prior_lsp_related, references/reasoning,references/models}
\bibliographystyle{icml2024}

%%%%%%%%%%%%%%%%%%%%%%%%%%%%%%%%%%%%%%%%%%%%%%%%%%%%%%%%%%%%%%%%%%%%%%%%%%%%%%%
%%%%%%%%%%%%%%%%%%%%%%%%%%%%%%%%%%%%%%%%%%%%%%%%%%%%%%%%%%%%%%%%%%%%%%%%%%%%%%%
% APPENDIX
%%%%%%%%%%%%%%%%%%%%%%%%%%%%%%%%%%%%%%%%%%%%%%%%%%%%%%%%%%%%%%%%%%%%%%%%%%%%%%%
%%%%%%%%%%%%%%%%%%%%%%%%%%%%%%%%%%%%%%%%%%%%%%%%%%%%%%%%%%%%%%%%%%%%%%%%%%%%%%%
\newpage
\appendix
\onecolumn
%\section{You \emph{can} have an appendix here.}

\input{sections/appendix}

\end{document}

%% file: commands.tex
\newcommand{\BasicPrompting}{\textsc{BasicPrompting}}
\newcommand{\Reflexion}{\textsc{Reflexion}}
\newcommand{\Repocoder}{\textsc{Repocoder}}
\newcommand{\NaiveRAG}{\textsc{NaiveRAG}}

\newcommand{\OurMethod}{\textsc{RRR}}

\newcommand{\Sketchy}{\textsc{Sketchy}}
\newcommand{\Descriptive}{\textsc{Detailed}}
\newcommand{\Detailed}{\textsc{Detailed}}
\newcommand{\expectedpass}{\textsc{EXPECTED\_TO\_PASS}}

\newcommand{\Java}{\textsc{Java}}
\newcommand{\Python}{\textsc{Python}}
\newcommand{\CSharp}{\textsc{CSharp}}

\newcommand{\datasetname}{RepoClassBench}

\newcommand{\greencheck}{{\color{green}\checkmark}} 
\newcommand{\redcross}{{\color{red}$\times$}}

\definecolor{bg}{rgb}{0.9, 0.9, 0.9}

\lstdefinestyle{PythonStyle}{
  language=Python,
  basicstyle=\ttfamily\small,
  backgroundcolor=\color{bg},
  frame=single,
  framesep=0pt,
  xleftmargin=1pt,
  xrightmargin=1pt,
  keepspaces=true
}

\newcommand{\pythoninline}[1]{
  \setlength{\fboxsep}{0.8pt}
  \raisebox{-1pt}{\colorbox{bg}{\lstinline[style=PythonStyle]!#1!}}
}

%% file: sections/introduction.tex
\section{Introduction}
Using LLMs to generate code has garnered significant attention in recent years for its potential to streamline software development processes by automatically translating natural language descriptions into executable code snippets. Several code-specific models, like CodeGen \citep{llms:codegen}, WizardCoder \citep{llms:wizardcoder}, CodeLlama \citep{llms:codellama}, StarCoder \citep{llms:starcoder},  DeepSeekCoder \citep{llms:deepseekcoder}, have been proposed to this end. 

While much of the focus in this domain has been on generating code units such as functions or statements, the specific task of generating classes has received comparatively less attention. Two of the most popular benchmarks HumanEval \citep{bench:humaneval} and MBPP \citep{bench:mbpp}, for instance, focus on function generation. While useful, the problems in these datasets are short and standalone, and existing works have been able to show good performance on these benchmarks. LATS \citep{reasoning:lats} for instance reports a 94.4\% accuracy on HumanEval, and 81.1\% accuracy on MBPP.

To address both of these issues, ClassEval \citep{bench:classeval} proposes a benchmark for class generation. The 100 classes ClassEval were handcrafted such that they contain inter-method dependencies, i.e. a method could reference another method in the same class. Using this dataset, they showed that, LLMs have a harder time generating code with these kind of dependencies than standalone functions of the kind present in HumanEval or MBPP.

While an important contribution, the problems proposed in ClassEval are still standalone when taking the class as a single unit. The only dependencies from outside the class are from well known libraries that the LLM is likely to have memorized. This narrow focus overlooks the complex dependencies that classes may have on other components within a codebase, presenting a gap in our understanding of code generation techniques' practical applicability. A much more useful problem is to consider the generation of a new class that depends on code from across a repository.

To address this gap, we create a dataset to explore the task of generating classes within the context of code repositories, where classes may interact with other code entities within a larger codebase. Specifically, we collect 130 Java classes from 10 repositories, 97 Python classes from 10 repositories and 60 C\# classes from a repository to create a benchmark called \textbf{\datasetname}. Each class is present in the context of a real-world repository and has dependencies on other parts of the repository. Additionally, we make sure that each class has corresponding test cases that pass on the ground truth, and ensure sufficient coverage.  

To be able to solve the problems in this dataset, the model has to both, understand the
functionality required from each method in the class and reason about how to use repository dependencies to achieve the same. We provide an evaluation of existing code-generation
techniques in this setting, and demonstrate their poor performance. Specifically, \BasicPrompting~either hallucinates identifiers or avoids the dependencies, \Reflexion~is
able to reason about the error, but does not have enough context to fix it, and RAG-based approaches are able to find similar snippets from across the repo but fail to bring in other kinds
of dependencies that are required by the class. Taking a step forward, we address the shortcoming of these methods, by proposing a novel method called \OurMethod~and show significant
gains. Specifically, \OurMethod~leverages existing programming language tools to retrieve precise
information from across the repository. With the injection of pointed repository context
through these tools, the model is able to fix the error observed during the feedback-reflection
stage.By bridging these gaps, our study seeks to contribute to a deeper understanding of LLMs' potential in generating classes within real-world coding scenarios, with implications for the development of more effective code generation techniques in the future. Our contributions are three-fold:

\begin{itemize}
\item  We contribute the first benchmark \datasetname~ for class-level code generation in realistic environment of an existing repository, with $130$ java classes (spanning 10 repositories), $97$ python classes (spanning 10 repositories) and $60$ C\# classes (spanning across a very large repository).
\item We propose a novel method "\OurMethod" that equips LLMs with static analysis tools to iteratively navigate \& reason about repository-level context in an agent-based framework, and compare \OurMethod~with existing methods.
\item We contribute 6 repository tools, based on our observations of common errors experienced by code agents in this setting. 
\end{itemize}

%% file: sections/related_works.tex
\section{Related Work}
% A list of possibly relevant work maintained on this \href{https://microsoftapc-my.sharepoint.com/personal/ariy_microsoft_com/_layouts/OneNote.aspx?id=%2Fpersonal%2Fariy_microsoft_com%2FDocuments%2FNotebooks%2FTMNT%20%28Tiered%20Model%20Ninja%20Team%29&wd=target%28Current%20Work%2FToolFramework.one%7C2F1D1A04-B755-428E-9A56-009A3FDA5EF8%2FReading%20List%20%28for%20RepoTools%5C%29%7C6592EF80-2F3D-4EDD-97C4-A34F4A44E63F%2F%29
% onenote:https://microsoftapc-my.sharepoint.com/personal/ariy_microsoft_com/Documents/Notebooks/TMNT%20(Tiered%20Model%20Ninja%20Team)/Current%20Work/ToolFramework.one#Reading%20List%20(for%20RepoTools)&section-id=\{2F1D1A04-B755-428E-9A56-009A3FDA5EF8\}\&page-id=\{6592EF80-2F3D-4EDD-97C4-A34F4A44E63F\}\&end}{OneNOTE page}.

LLMs have seen wide success on various coding tasks. Many benchmarks have been created to assess their performance. CoNaLA \citep{bench:CoNaLA}, consisting of 500 examples is a statement-level benchmark where the target of each example contains one statement. HumanEval \citep{bench:humaneval} and MBPP \citep{bench:mbpp} are two widely used datasets, for function level code-generation, consisting of 164 and 974 tasks respectively. At the class-level, ClassEval \citep{bench:classeval} has been proposed with 100 class generation problems, where the input is the class skeleton. However, these are all independent code-generation problems. Although ClassEval includes inter-method dependencies, they are all present within the same class. The external references come from well-known libraries that the LLM is likely to have memorized. In real world repositories, code includes complex inter-dependencies from other files in the repository. RepoBench \citep{bench:repobench}, CoderEval \citep{bench:coder_eval} and MGD \citep{bench:mgd} are attempts to move closer to this setting, and show that existing models perform much better on the standalone setting than the non-standalone setting. However they explore line and function level tasks in the context of a repository, whereas \datasetname~ explores the generation of non-standalone classes within the context of a repository. There are two aspects to solving our dataset, retrieving the right context, and reasoning to generate the code.

\textbf{Reasoning:}
To improve the generation of LLMs, various iterative refinement techniques have been proposed. Self-refine \citep{reasoning:self_refine} attempts to use the LLM as it's own critic and produces successively better outputs. Reflexion \citep{reasoning:reflexion} incorporates test-case feedback while generating the reflection on its output. LATS \citep{reasoning:lats} uses the LLM as an agent to explore a tree of solutions, using compiler and test feedback as observations.

\textbf{Retrieval:}
While reasoning-enhanced methods, in themselves, may be useful for standalone generations, they are not sufficient when external context is needed. This is especially true, when the context consists of private data, unseen during pretraining. Under this paradigm Retrieval-Augmented-Generation methods like REALM \citep{cr:realm}, ATLAS \citep{cr:atlas}, RetGen \citep{cr:retgen}, FLARE \citep{cr:flare} retrieve relevant context, usually by considering snippets with the highest similarity score with the query. Similarly, in the code setting, RLPG \citep{cr:rlpg} trains a model to predict the relevant context source, but relies on there being a "hole" in the code, whereas there is no such hole in the NL to new class setting. Additionally the RLPG model was trained for Java, whereas for the other languages new models would need to be trained. This adds additional cost of constructing new training data and the actual training of new models. RepoCoder \citep{cr:repocoder} has been proposed to perform iterative retrieval and generation. While such similarity based RAG methods can retrieve "similar" context, they fail to effectively retrieve "dependency" context. Further discussion can be found in RQ2. 

In our method, we leverage repository-level tools to allow the LLM explore the repository, as an alternative retrieval mechanism, in addition to using test-case feedback.  This is along the lines of several works that have explored equipping the LLM with tools like ReACT \citep{tools:react} and ToolFormer \citep{tools:toolformer}. However to our knowledge, this is the first work that curates tools-specifically for repository-exploration.

Hence, we propose a benchmark that addresses the problem of class generation in the context of a repository, address a gap in the span of existing benchmarks, and also propose a novel method that integrates retrieval and reasoning, mitigating the shortcomings of existing methods.

%% file: sections/data_generation_methodology.tex
\section{Dataset: \datasetname}
\datasetname{} is a benchmark featuring repositories from Github across languages: \Java, \Python~and \CSharp. The task is to synthesize a complete class within a repository based on a natural language description, utilizing the context from other files within the same repository.

Current benchmarks face two primary limitations: (1) they \citep{bench:classeval} typically focus on generating small localized code snippets, which do not accurately represent the complex tasks software engineers encounter, often requiring a comprehensive understanding of the entire codebase; (2) they \citep{bench:repobench} rely on metrics such as exact-match or cosine-similarity to the ground truth for evaluation, rather than assessing the functionality of the generated code through test cases. We mitigate these issues by designing a benchmark where every task corresponds to a class-generation problem where the LLM needs to synthesize the class based on the natural language specification of the class. We ensure that every class in our benchmark makes use of external references in the repository and is covered under test cases. 

\subsection{Benchmark Construction} 
\textbf{Stage 1 - Shortlisting repositories:}
Our benchmark includes repositories both before and after the cutoff-date of the models we evaluate on. For \Java~we start with repositories considered in the MGD \citep{bench:mgd} dataset. For Python, we adapt the popular benchmark SWEBench \citep{swebench} and also shortlist popular repositories which were first created on GitHub after Sept 2021. For \CSharp~we consider the singular large repository \textit{StabilityMatrix}. We filter out those repositories which we are unable to build and run. (Details in \ref{sec:repository-selection})

\textbf{Stage 2 - Shortlisting classes:} Within each repository, we identify all classes that pass the existing test cases. We retain only those classes that (a) reference other parts of the repository within their body, and (b) have methods covered by test cases. Additionally, we limit our selection to classes defined in the global namespace. (Details in \ref{sec:class_shortlist}) 

\textbf{Stage 3 - Dataset paraphrasing:} For repositories available before the LLMs' training data cutoff, we undertake a paraphrasing initiative, altering the names of most symbols to prevent models from completing tasks through mere memorization. (Details in \ref{sec:dataset-paraphrasing})
 
\textbf{Stage 4 - Generating natural language specification:} We break the information within each class into varying levels of granularity and record it as metadata. The complete metadata fields are listed in Table \ref{table:metadata_components}. Methods are categorized by three information levels: (1) Signature, detailing input and output types; (2) Docstring, providing a high-level function description; (3) Body, outlining full implementation and logic, including external references. We prompt GPT-4 to generate the natural language description of the class by providing it varying granularity of information extracted as a subset of the metadata (refer to Table \ref{table:metadata_components}). Hence, two types of natural language description in our dataset are:-
\begin{enumerate}
    \item \Detailed : This includes details from the entire class body (excluding imports) and prompts GPT-4 to create an NL description.
    \item \Sketchy:  This omits method bodies from the prompt, leading GPT-4 to generate an NL description without low-level implementation specifics or explicit external references.
\end{enumerate}
In the \Sketchy{} setting, since GPT-4 does not receive the method bodies, the resulting natural language (NL) descriptions lack detailed implementation specifics and explicit mentions of the external references used during the method's development. Consequently, the \Sketchy{} NL descriptions present a higher level of difficulty compared to the \Descriptive{} versions. To foster community engagement and further research, we make the metadata used for constructing these prompts publicly available. This allows others to create NL descriptions with varying degrees of specificity and ambiguity to challenge the models' capabilities. Example of the difference in prompts to GPT-4 for them can be found in Prompt \ref{diffs:example1}.

Some \textbf{statistics} about our dataset can be found in Table \ref{tab:dataset_high_level_statistics}. Distribution of tasks across different repositories can be found in: Figures \ref{fig:label1} and \ref{fig:label2}.

%% file: sections/method.tex
\section{Method}
To address the challenges presented by our benchmark, we propose Retrieve-Repotools-Reflect (\OurMethod), an innovative method that enhances Large Language Models (LLMs) with static analysis tools. This approach enables the LLMs to iteratively explore and understand the context of a code repository through an agent-based framework. \OurMethod~leverages repository navigation and reasoning capabilities to effectively synthesize code that aligns with the broader structure and dependencies of the repository.

\input{sections/appendix/algorithm}

\subsection{Phases of \OurMethod}
The procedural framework of \OurMethod~is illustrated visually in Figure \ref{fig:methodfigure} and outlined algorithmicaly in Algorithm \ref{algo:main_algorithm}.
During the initial generation phase, the LLM $\mathcal{M}$ makes an initial "guess" $y_{1}$ based on the class description $x$ and output from invocations of the \textit{independent tool} $t_{0}$: $y_{1}=\mathcal{M}(x,t_{0})$. 
Given the limited information available at this stage, the LLM may resort to hallucinating identifiers and other code-structures. (Prompt in \ref{prompts:initial})

The oracle call entails passing the generated code $y_{i}$ to the oracle $\mathcal{O}$, to receive \textit{oracle feedback} $fb_i$, $fb_{i}=\mathcal{O}(y_{i})$. If the attempt exceeds the maximum number of oracle calls or successfully passes all test cases, the loop terminates and returns $y_i$. Otherwise, the oracle feedback errors $fb_i$ are utilized by the LLM agent in subsequent phases to refine its generation.

While the oracle feedback identifies problems in the code, it lacks guidance on error resolution. To address this, the LLM requires repository context.  This context is provided through carefully curated tools, allowing the LLM to explore the repository and retrieve relevant information. Based on the class description $x$,  current generation $y_{i}$ and feedback $fb_{i}$, the model generates \textit{a set of tool calls} $\mathbb{T}_{i}$: $\mathbb{T}_{i}=\mathcal{M}(x,y_{i},fb_{i})$. The executor $\mathcal{E}$ takes these tool calls and produces outputs $t_{i}$: $t_{i}=\mathcal{E}(\mathbb{T}_{i})$. (Prompt in \ref{prompts:toolcall})

Based on the feedback from the oracle $fb_{i}$ and tool outputs $t_{i}$, the LLM generates a reflection $r_{i}$ on the encountered errors and necessary actions to rectify them, using hints from the tool outputs $t_{dependent}$. $r_i=\mathcal{M}(x,y_{i},fb_{i},t_{i})$ This reflection serves as a hint for the subsequent stage. (Prompt in \ref{prompts:reflection})

In the improved generation phase, leveraging the last attempt's $y_{i}$, oracle feedback $fb_{i}$, tool outputs $t_{i}$, and reflection $r_{i}$, the LLM makes another attempt at code generation $y_{i+1}$. $y_{i+1}=\mathcal{M}(x,y_{i},fb_{i},t_{i},r_i)$ (Prompt in \ref{prompts:codegen})

After the improved generation, the attempt gets passed back to the "Oracle call" phase and the loop continues.

\input{tables_used/tool_menu}
\subsection{Tools}
In \OurMethod, tools are categorized as either independent or dependent based on their need for reasoning. \textbf{Independent tools} operate without considering the current state of the \OurMethod~loop and are automatically invoked during the initial generation phase. Our suite includes a single independent tool, `get\_related\_snippets`. On the other hand, tools requiring reasoning over the current state of the \OurMethod~loop are classified as \textbf{dependent tools}. Our dependent toolset contains get\_imports, get\_class\_info, get\_signature, get\_method\_body and get\_relevant\_code. More information about the tools can be found in Table \ref{tab:tools}.

%% file: sections/appendix/algorithm.tex
%\section{Algorithm}

\begin{algorithm}[H]
\caption{\OurMethod}
\label{algo:main_algorithm}
\begin{algorithmic}[1]
\Procedure{\OurMethod}{$x$: Natural language description, $MAX\_CALLS$: Maximum iterations,  $\mathcal{O}$: Oracle function, $\mathcal{M}$: Language Model Agent, $\mathcal{E}$: Tool executor}
\State $y_1 = \mathcal{M}(x, t_0)$  \Comment{Initial guess using NL description, independent tool outputs}
\For{$i = 1$ \textbf{to} $MAX\_CALLS$} \Comment{Iterate over generation attempts}
    \State $fb_i = \mathcal{O}(y_i)$  \Comment{Get feedback from oracle}
    \If {all test cases pass in $fb_i$}
        \State \textbf{return} $y_i$
    \EndIf    
    \State $\mathbb{T}_i = \mathcal{M}(x, y_i, fb_i)$  \Comment{Generate tool calls based on the error}
    \State $t_i = \mathcal{E}(\mathbb{T}_i)$  \Comment{Execute tool calls}
    \State $r_i = \mathcal{M}(x, y_i, fb_i, t_i)$  \Comment{Generate reflection}
    \State $y_{i+1} = \mathcal{M}(x, y_i, fb_i, t_i, r_i)$  \Comment{Generate improved code}
\EndFor
\State \textbf{return} $y_{i+1}$ \Comment{Return the latest generated code}
\EndProcedure
\end{algorithmic}
\end{algorithm}

%% file: tables_used/tool_menu.tex
\begin{table}[htbp]

\small
\renewcommand{\arraystretch}{1.5} 
\begin{tabular}{p{2cm}p{11cm}}
\toprule
\textbf{Tool} & \textbf{Description} \\
\midrule
\textbf{get\_related \_snippets} & \textit{Type: Independent.} Segments the repository into snippets and returns the top 5 snippets based on cosine similarity with the class description. \\

\textbf{get\_imports} & \textit{Type: Dependent.} Suggests imports for all the undefined symbols in the current generation, scanning the repository for potential source files defining the symbol and recommending import statements. \textit{Input args:} No input\\

\textbf{get\_class\_info} & \textit{Type: Dependent.} Locates the class definition in the repository and gathers information about its members, including inherited members, providing detailed information about each member. \textit{Input args:} class name \\

\textbf{get\_signature} & \textit{Type: Dependent.} Returns the signature of the requested method, displaying signatures of all methods with the same name if they exist in the same class. \textit{Input args:} class name, method name \\

\textbf{get\_method \_body} & \textit{Type: Dependent.} Returns the method definition of the requested method, truncating the output if it is too large, and showing the definition for each method with the same name if they exist. \textit{Input args:} class name, method name (where class name is the class of which the method is a member. Class name is left as None for global methods.) \\

\textbf{get\_relevant \_code} & \textit{Type: Dependent.} Allows specific queries to retrieve code structures using embedding similarity scores, returning the top 3 structures based on cosine similarity using UnixCoder embeddings. \textit{Input args:} natural language query \\
\bottomrule
\end{tabular}
\caption{Table containing descriptions of the tools used in \OurMethod. The Type indicates whether reasoning is required (dependent) or not (independent) for the invocation. }
\label{tab:tools}
\end{table}

%% file: sections/experimental_results.tex
\section{Experimental Results}

\input{tables_used/baseline_summary}

\subsection{Baselines}

Apart from \OurMethod, we test other important baselines (summarized in Table \ref{tab:approaches}) on our newly constructed benchmark. In \BasicPrompting~the LLM is expected to generate code solely based on the Natural Language Description. In \NaiveRAG, inputs include the Natural Language Description and and top-snippets retrieved from repository when queried using the Natural Language Description. \Reflexion~incorporates Oracle feedback to iteratively improve the generation. We also use \Repocoder, where the initial generation uses snippets retrieved using the Natural Language Description as the query, and subsequent iterations use snippets retrieved using the previous code-generation as the query. Summary of the features of the different baselines can be found in Table \ref{tab:approaches}.

\subsection{Metrics}
For each task in our benchmark we use three metrics to measure performance.  
\begin{itemize}
\item \textbf{Pass@K} indicates the percentage of tasks where at least one correctly generated solution (passing all test cases) is found among the top K samples generated by the Language Model (LLM) \citep{bench:humaneval}. We set K=1 and report the mean and standard deviation of this metric computed over 6 independent runs. For completeness, in RQ 7, we also measure Pass@ 1,2,3.
\item We also use \textbf{TR (Test Rate)} which measures the mean of the fraction of test cases passed for all generations across all tasks. 
\item Finally, for \Java~and \CSharp, since we have access to a compiler, we also measure \textbf{CR, or the compilation rate} which measures the percentage of tasks for which the LLM  generated code that successfully compiled.
\end{itemize}

\subsection{Research Questions}
Through our experiments, we aim to answer the following research questions (RQs): \hyperref[rq:performance_comparison]{RQ1} - How does \OurMethod~perform compared to the baselines in both \Detailed~and \Sketchy~settings? \hyperref[rq:cousinsrq]{RQ2} - Where do similarity-based retrieval methods fail? \hyperref[rq:tc_feedback]{RQ3} - What is the impact of test feedback on performance? \hyperref[rq:failureanalysisrq]{RQ4} - What are the challenges faced by \OurMethod? Additional RQs 5-9, discussing the role of different tools, the impact of sampling, and the number of feedback iterations, are available in Appendix, Section \ref{sec:additional-rqs}.

\input{sections/research_questions/rq1}

\input{sections/research_questions/rq2}

\input{sections/research_questions/rq3}

\input{sections/research_questions/rq4}

%% file: tables_used/baseline_summary.tex
\begin{table}[h]
    \centering
    \begin{tabular}{lccccc}
    \toprule
    \textbf{Approach} & 
    \textbf{External Context} &
    \textbf{Testcase feedback} &
    \textbf{Multiple} &
    \textbf{Tools access}  \\
    & \textbf{shown} & \textbf{for refinement} & \textbf{iterations} &  \\
    \midrule
    \BasicPrompting & \redcross & \redcross & \redcross & \redcross \\
    \Reflexion & \redcross & \greencheck & \greencheck & \redcross \\
    \NaiveRAG  & \greencheck & \redcross & \redcross & \redcross \\
    \Repocoder & \greencheck & \redcross & \greencheck & \redcross \\
    \OurMethod  & \greencheck & \greencheck & \greencheck & \greencheck \\
    \bottomrule    
    \end{tabular}
\caption{A breakdown of the different approaches compared in this work. \OurMethod~ticks all the boxes.}
\label{tab:approaches}
\end{table}

%% file: sections/research_questions/rq1.tex
\subsubsection{RQ1 - Comparison of \OurMethod~with baselines}
\label{rq:performance_comparison}

{
\tabcolsep=0.075cm
\begin{table}[ht]
\small
\centering
\begin{tabular}{lcccccccccccc}
\toprule
\bf Method & \multicolumn{3}{c}{\bf \Java} & \multicolumn{3}{c}{\bf \CSharp} & \multicolumn{2}{c}{\bf \Python} \\
\cmidrule(lr){2-4}
\cmidrule(lr){5-7}
\cmidrule(lr){8-9}
& \bf P@1 &\bf TR & \bf CR & \bf P@1 &\bf TR & \bf CR & \bf P@1 &\bf TR  \\
\midrule
\BasicPrompting & 2.18 (0.29) & 2.18 (0.29) & 3.08 (0.44) & 0.00 (0.00) & 0.00 (0.00) & 0.00 (0.00) & 1.2 (0.38) & 3.1 (0.72) \\ 
\Reflexion & 18.59 (1.36) & 20.28 (0.82) & 23.46 (1.82) & 2.82 (1.6) & 3.11 (2.06) & 3.39 (1.96) & 8.42 (0.7) & 12.93 (0.8)  \\ 
\NaiveRAG & 14.23 (0.38) & 14.56 (0.24) & 17.69 (0.99) & 1.69 (0.00) & 1.69 (0.00) & 3.11 (0.63) & 17.01 (1.54)  & 17.73 (1.45) \\ 
\Repocoder & 54.74 (1.03) & 58.12 (1.39) & 62.44 (1.21) & 1.98 (0.63) & 1.98 (0.63) & 1.98 (0.63) & 23.54 (2.62) & 26.83 (1.66) \\ 
\OurMethod & \textbf{77.56 (2.06)} & \textbf{84.27 (1.47)} & \textbf{92.56 (1.23)} & \textbf{33.9 (4.27)} & \textbf{40.96 (3.45)} & \textbf{42.37 (3.79)} & \textbf{29.38 (1.29)} & \textbf{35.69 (1.14)} \\
\bottomrule
\end{tabular}
\caption{Performance numbers expressed in percentage, for the baselines and \OurMethod~on the \textbf{\Detailed~version} of the dataset. P@1 represents the Pass@(1,1) metric, TR is the mean test-pass rate across all tasks, and CR is the mean compilation rate across tasks. Since \Python~does not have a compiler, this column is skipped. The numbers in the bracket represent the standard deviation over 6 generations. \OurMethod~performs much better than the baselines.}
\label{tab:rq1_fully_specified}
\end{table}
 }

{
\tabcolsep=0.075cm
\begin{table}[ht]
\small
\begin{center}
\begin{tabular}{lccccccccc}
\toprule
\bf Method & \multicolumn{3}{c}{\bf \Java} & \multicolumn{3}{c}{\bf \CSharp} & \multicolumn{2}{c}{\bf \Python} \\
\cmidrule(lr){2-4}
\cmidrule(lr){5-7}
\cmidrule(lr){8-9}
& \bf P@1 &\bf TR & \bf CR & \bf P@1 &\bf TR & \bf CR & \bf P@1 &\bf TR  \\
\midrule
\BasicPrompting & 0.9 (0.29) & 1.59 (0.29) & 3.59 (0.57) & 0.00 (0.00) & 0.00 (0.00) & 0.00 (0.00) & 0.69 (0.76) & 3.33 (1.08) \\ 
\Reflexion & 11.03 (0.57) & 12.89 (1.23) & 17.95 (1.31) & 3.67 (0.63) & 3.67 (0.63) & 3.67 (0.63) & 5.67 (0.98) & 13.41 (1.23) \\ 
\NaiveRAG & 9.23 (0.63) & 10.97 (0.51) & 15.13 (1.58) & 0.00 (0.00) & 0.00 (0.00) & 1.69 (0.00) & 11 (1.6) & 13.04 (1.59) \\ 
\Repocoder & 42.05 (0.85) & 50.37 (0.64) & 60.38 (1.31) & 0.28 (0.63) & 0.28 (0.63) & 0.28 (0.63) & 18.04 (0.78)  & 23.07 (1.0) \\ 
\OurMethod & \textbf{70.64 (1.74)} & \textbf{77.94 (0.57)} & \textbf{88.21 (1.06)} & \textbf{23.73 (2.59)} & \textbf{27.68 (1.26)} & \textbf{29.94 (0.8)} & \textbf{19.24 (1.28)} & \textbf{29.7 (0.85)} \\

\bottomrule
\end{tabular}\end{center}
\caption{Performance numbers expressed in percentage, for the baselines and \OurMethod~on the \textbf{\Sketchy~version} of the dataset.  Since \Python~does not have a compiler, the CR column is skipped. The numbers in the bracket represent the standard deviation over 6 generations. \OurMethod~performs much better than the baselines.}
\label{tab:rq1_dereferenced_table}
\end{table}
}

We compared the performance of \OurMethod with baseline methods using GPT-4 on \Java, \Python, and \CSharp, as shown in Tables \ref{tab:rq1_fully_specified} and \ref{tab:rq1_dereferenced_table}. \OurMethod~ consistently surpassed the baselines across all evaluation metrics. Without feedback or context, \BasicPrompting~ had the lowest performance, with very few generated code snippets passing test cases. Although \Reflexion~ showed some improvement with oracle feedback, it still performed poorly due to the absence of repository context, often resulting in the generation of incorrect identifiers and underutilization of repository resources.

To add the repository context one might consider dumping the entire repository in the prompt. However, the token count in the repositories considered can exceed 50k, surpassing LLM context windows, and dumping entire repositories into prompts is impractical. To tackle these issues, methods that employ retrieval can be used. There's a noticeable performance jump from \Reflexion~to \NaiveRAG, further improved with \Repocoder, due to more relevant retrieved snippets.  While \Repocoder is the best performing baseline, it has two major drawbacks. It does not utilize oracle feedback and its retrieval mechanism focuses on fetching "similar" lines of code rather than dependencies, thus omitting vital information. This issue is examined in greater depth in \hyperref[rq:cousinsrq]{RQ2}. Conversely, \OurMethod~retrieves dependency context, combining repository context and oracle feedback intelligently. It queries specific repository information to address oracle feedback, consistently outperforming baselines across languages and metrics. Still, there are cases where \OurMethod~fails test cases, which we analyze in \hyperref[rq:failureanalysisrq]{RQ4}.

It is also observable that \Java's performance numbers are greater than \CSharp~and \Python. This can be attributed to the sizes of classes in the dataset. Referring to Table. 12, it can be observed that the average length of the \Java~classes is  2080 characters and the average length of the \Python~and \CSharp~classes are 4663.76 and 4283.00 characters respectively. Thus, on an average \Python~and \CSharp~classes tend to be more than double the size of \Java~classes. Since the model is prone to making more errors with longer classes, this difference in performance can be observed.

%% file: sections/research_questions/rq2.tex
\subsubsection{RQ2 - Contribution of similarity-based RAG}
\label{rq:cousinsrq}

\begin{table}[ht]
\small
\begin{center}
\begin{tabular}{lccccccccccccccccc}
\toprule
\bf Method & \multicolumn{3}{c}{\bf \Java - \Detailed} & \multicolumn{3}{c}{\bf \Java - \Sketchy} & \multicolumn{3}{c}{\bf \CSharp - \Detailed} & \multicolumn{3}{c}{\bf \CSharp - \Sketchy} \\
\cmidrule(lr){2-4}
\cmidrule(lr){5-7}
\cmidrule(lr){8-10}
\cmidrule(lr){11-13}
& \bf P@1 &\bf TR & \bf CR & \bf P@1 &\bf TR & \bf CR & \bf P@1 &\bf TR & \bf CR & \bf P@1 &\bf TR & \bf CR \\
\midrule
\BasicPrompting & 2.31 & 2.31 & 2.31 & 0.77 & 1.46 & 2.31 & 0.00 & 0.00 & 0.00 & 0.00 & 0.00 & 0.00 \\ 
\Reflexion & 20.77 & 23.36 & 27.69 & 11.54 & 13.34 & 20.77 & 3.39 & 3.39 & 3.39 & 3.39 & 3.39 & 3.39 \\ 
\NaiveRAG & 11.54 & 11.54 & 14.62 & 7.69 & 9.19 & 13.08 & 0.00 & 0.00 & 0.00 & 0.00 & 0.00 & 0.00 \\ 
\Repocoder & 42.31 & 44.99 & 48.46 & 29.23 & 37.78 & 47.69 & 1.69 & 1.69 & 1.69 & 0.00 & 0.00 & 0.00 \\ 
\OurMethod & 72.31 & 77.75 & 84.62 & 61.54 & 68.95 & 83.85 & 33.90 & 42.37 & 42.37 & 20.34 & 22.03 & 28.81 \\

\bottomrule
\end{tabular}
\end{center}
\caption{Performance numbers expressed in percentage, for the baselines and \OurMethod, after removing the "Relatives" from the \Detailed~and \Sketchy~versions of the Java dataset. While all retrieval-based methods suffer, \OurMethod~does not suffer as much as \Repocoder.}
\label{tab:rq2}
\end{table}

In this benchmark, typical GitHub repositories feature numerous highly similar classes. RAG-based methods outperform \BasicPrompting~ and \Reflexion~ by leveraging these similarities. Yet, it's critical to differentiate between "dependency context" and "similarity context." Dependency context pertains to repository information about used code structures, while similarity context merely identifies similar code, which isn't always available.

To show that \Repocoder's advantages are mainly due to "similar" snippets, we removed all class "relatives" defined as descendants of the grandparent class, excluding the immediate parent and the class itself. These relatives are often similar to the target class and are retrieved by \Repocoder. After their removal and re-evaluation against baselines (see Table \ref{tab:rq2}), \Repocoder's performance drops in both \Detailed~ and \Sketchy~ settings. In contrast, \OurMethod~ is less affected, indicating its reliance on "dependency context" rather than "similarity context" for code generation.

%% file: sections/research_questions/rq3.tex
\subsubsection{RQ3 - Importance of test feedback}
\label{rq:tc_feedback}

{
\begin{table}[ht]
\small
\begin{center}
\begin{tabular}{lccccccccccccccccc}
\toprule
\bf Method & \multicolumn{3}{c}{\bf \Java - \Detailed} & \multicolumn{3}{c}{\bf \Java - \Sketchy} & \multicolumn{3}{c}{\bf \CSharp - \Detailed} & \multicolumn{3}{c}{\bf \CSharp - \Sketchy} \\
\cmidrule(lr){2-4}
\cmidrule(lr){5-7}
\cmidrule(lr){8-10}
\cmidrule(lr){11-13}
& \bf P@1 &\bf TR & \bf CR & \bf P@1 &\bf TR & \bf CR & \bf P@1 &\bf TR & \bf CR & \bf P@1 &\bf TR & \bf CR \\
\midrule
\BasicPrompting & 2.18 & 2.18 & 3.08 & 0.90 & 1.59 & 3.59 & 0.00 & 0.00 & 0.00 & 0.00 & 0.00 & 0.00 \\ 
\Reflexion & 13.46 & 16.63 & 27.18 & 6.41 & 9.89 & 18.97 & 2.54 & 2.82 & 3.39 & 3.39 & 3.39 & 4.24 \\ 
\NaiveRAG & 14.23 & 14.56 & 17.69 & 9.23 & 10.97 & 15.13 & 1.69 & 1.69 & 3.11 & 0.00 & 0.00 & 1.69 \\ 
\Repocoder & 53.97 & 57.68 & 63.72 & 40.51 & 49.24 & 61.92 & 1.98 & 1.98 & 3.39 & 0.28 & 0.28 & 1.98 \\ 
\OurMethod & 65.13 & 76.90 & 93.59 & 53.97 & 68.09 & 88.85 & 31.64 & 39.55 & 43.22 & 23.45 & 27.12 & 30.23 \\ 

\bottomrule
\end{tabular}
\end{center}
\caption{Performance numbers expressed in percentage, for the baselines and \OurMethod, terminating the generation immediately after the compilation succeeds, on the \Detailed~and \Sketchy~versions of the Java dataset. There is a marginal decrease in performance, indicating that most functional requirements can be met simply by using the compiler as the oracle.}
\label{tab:rq3}

\end{table}
}
Oracle feedback comprises both compiler feedback and testcase feedback—the former indicating compilation success or errors, and the latter providing information on which test cases the current code fails. To investigate the impact of testcase feedback on different methods, we restricted oracle feedback to only compiler feedback for \Java~ and \CSharp. Table \ref{tab:rq3} shows that baselines like \BasicPrompting, \NaiveRAG, and \Repocoder~ are unaffected by the absence of testcase feedback. Methods that leverage test feedback exhibit a slight decline in performance when restricted to compiler feedback, yet they still function effectively. Code that compiles and adheres to functional descriptions generally passes test cases, which primarily assess functional requirements. While test feedback can be useful for resolving ambiguities, the LLM often performs well with just compiler feedback.

%% file: sections/research_questions/rq4.tex
\subsubsection{RQ4 - Success and failure case analysis}
\label{rq:failureanalysisrq}
{
\tabcolsep=0.15cm
\begin{table}[!ht]
\small
\begin{tabular}{lccccc}
\toprule
\textbf{Error Type} & \multicolumn{2}{c}{\textbf{\Java}} & \multicolumn{2}{c}{\textbf{\CSharp}} \\
\cmidrule(lr){2-3}
\cmidrule(lr){4-5}
 & \textbf{Detailed} & \textbf{Sketchy} & \textbf{Detailed} & \textbf{Sketchy} \\
\midrule
Reasoning Errors & 50\% & 20\% & 40\% & 20\% \\
Functional Ambiguity & 50\% & 80\% & 60\% & 80\% \\
\bottomrule
\end{tabular}
\caption{Analyzing failure causes across a sample of 20 tasks from the \Java~and \CSharp~datasets, errors are categorized as reasoning-related (in tool retrieval or code generation) or functional ambiguity-related. The table shows the percentage contribution of each error type to failure cases. In the \Detailed~dataset, reasoning errors dominate, while in the \Sketchy~version, functional ambiguity-related errors increase.}
\label{tab:failure_analysis}
\end{table}
}
This section investigates instances where the different approaches failed to pass test cases, identifying potential contributing factors. Notably, errors weren't due to information access limitations through tools; there was always a tool for repository information retrieval. Our analysis focuses on categorizing error types to guide future investigations for mitigation strategies. 
Distinct error patterns emerged upon examination, broadly categorized as reasoning errors or functional ambiguity errors. Reasoning errors occur during tool retrieval or code generation, where the LLM fails to interpret or apply information correctly. Functional ambiguity errors arise when the LLM misinterprets terse natural language descriptions, leading to multiple interpretations or missing information. Table \ref{tab:failure_analysis}, a qualitative analysis of 20 failure cases, shows reasoning errors were higher in the \Detailed~setting, than in the sketchy setting. Additionally, the LLM struggles with lengthy textual inputs, with extended class length correlating significantly with decreased efficacy. Over the detailed \Java~and \CSharp~datasets, test performance and class length had Spearman correlations of -0.50 and -0.57 respectively, highlighting the challenge of reasoning over extensive texts. Identifying these failure cases sheds light on the dataset's role in understanding LLM capabilities and limitations. By pinpointing error patterns and correlating them with variables like class length, our analysis sets the stage for future research on enhancing language model robustness and efficacy.

%% file: sections/conclusion.tex
\section{Conclusion and Future Work}
\datasetname~offers a novel setting with unique challenges, that require reasoning over the repository. We highlight limitations in previous methods relying on similarity-based retrieval, and propose a solution leveraging both traditional embedding-based retrieval and static analysis tools. Through iterative refinement guided by tool outputs and oracle feedback, \OurMethod~outperforms existing baselines. However, utilizing tools in this manner has its caveats. Firstly, the context size limitation of the LLM prevents us from propagating history across iterations, and therefore, information required in later iterations needs to be re-acquired each time. Additionally, longer prompts exacerbate LLM failures and reasoning issues. Moreover, relying on tool outputs exposes us to vulnerabilities inherited from the base static analysis tools. These challenges warrant attention in future research. Nonetheless, our work underscores the potential of integrating tools with \OurMethod~to markedly enhance performance over baselines.

%% file: sections/appendix.tex
% \section{Baselines}
% Table \ref{tab:approaches} breaks down the different approaches compared in this work into their components. \OurMethod~utilizes all the components and thus is able to outperform the baselines.

\input{sections/appendix/experimental_setup}

\input{sections/appendix/additional_analysis}

\input{sections/appendix/dataset_generation}

\input{sections/appendix/tool_descriptions}

\begin{figure}[h]
    \centering
    \includegraphics[scale=0.7]{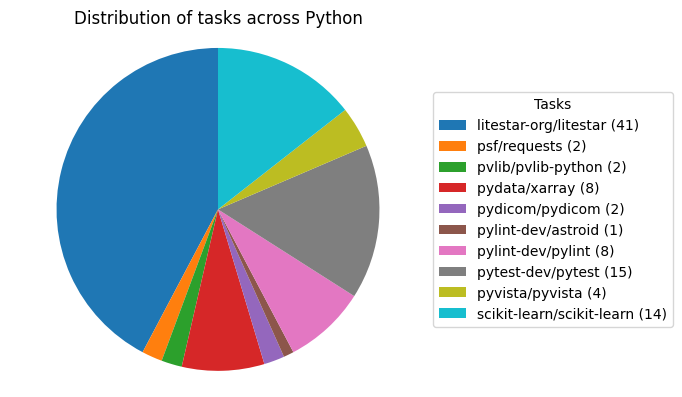}
     \caption{Distribution of the tasks across the various repositories in the Python dataset.}
    \label{fig:label1}     
\end{figure}

\begin{figure}[h]
    \centering
    \includegraphics[scale=0.7]{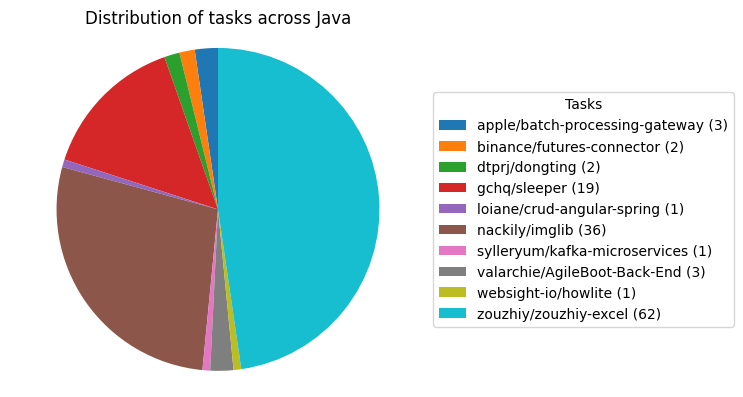}
    \caption{Distribution of the tasks across the various repositories in the Java dataset.}
    \label{fig:label2}    
\end{figure}

\section{An example showcasing the \OurMethod~workflow}
\label{eg:heading}
\input{motivating_examples/example_1/run_1}

\input{sections/appendix/prompt_info}

%% file: sections/appendix/experimental_setup.tex
\section{Experimental Setup}

We utilized the GPT-4 model, through the Azure OpenAI endpoint, configured with a temperature of 0.2, for all the experiments unless explicitly mentioned otherwise. Default values were maintained for all other parameters. All experiments were conducted on a machine with an AMD EPYC 7V13 64-Core Processor running at 2.45GHz,  216GB of RAM, and an NVIDIA A100 80GB GPU. The experiments were executed using Python 3.11.8 and PyTorch 2.0.1.

%% file: sections/appendix/additional_analysis.tex
\section{Additional Research Questions}
\label{sec:additional-rqs}

\input{sections/research_questions/rq5}

\input{sections/research_questions/rq6}

\input{sections/research_questions/rq7}

\input{sections/research_questions/rq8}

\input{sections/research_questions/rq9}

%% file: sections/research_questions/rq5.tex
\subsubsection{RQ5 - Tool statistics}

\begin{table}[ht]
\centering
\begin{tabular}{lccccccccc}
\toprule
\bf Method & \bf imports & \bf class\_info & \bf relevant\_code & \bf method\_body & \bf signature \\
\midrule
\Detailed & 28.76 & 34.44 & 26.19 & 4.02 & 6.59 \\ 
\Sketchy & 27.84 & 34.01 & 26.13 & 4.97 & 7.03 \\ 
\Detailed$_{NoRelatives} $ & 27.92 & 34.94 & 25.31 & 4.81 & 7.02 \\ 
\Sketchy$_{NoRelatives}$ & 27.37 & 33.54 & 26.63 & 5.80 & 6.54 \\ 

\bottomrule
\end{tabular}
\caption{The frequency distribution of tool calls, expressed in percentage, for the various configurations of the Java dataset. The LLM is biased to utilize the tools shown in the few-shot example (get\_imports, get\_class\_info, get\_relevant\_code for \Java), and makes superfluous calls using them, adding noise to the true frequency distribution. There is however, an expected increase in the frequency of certain tools like get\_relevant\_code, moving from \Detailed~to \Sketchy~and Relatives to No Relatives.}
\label{tab:rq4_tool_frequency}
\end{table}

While it is hard to evaluate exactly how much each tool contributed to the the success of our method, a proxy for this measurement could be to analyze the frequency with which  each tool is called, in all the cases where the generation passed all test cases. Thus, in Table \ref{tab:rq4_tool_frequency} we count the number of times each tool was called, for all the successful generations, and express these counts as percentages. However, we notice an LLM-related idiosyncrasy, where the model is strongly biased to using the tools shown to it in the few-shot examples. Even in cases where certain tools are not required, it forces itself to create a reason for using it and makes superfluous calls, adding noise to the frequency count. To illustrate this point more clearly, consider Table Table \ref{tab:rq4_tool_presence}. Here, each cell represents the percentage of tasks, across which the corresponding tool was called at least once (assuming at least one round of tool-invocation happened). As visible from the table, the tools present in the few-shot example (get\_imports, get\_class\_info, get\_relevant\_code for \Java), were called in every example. The other tools get called much less frequently. This motivates careful selection of the few-shot examples, to choose tools which would be required to solve the most frequently observed errors.

\begin{table}[ht]
\centering
\begin{tabular}{lccccccccc}
\toprule
\bf Method & \bf imports & \bf class\_info & \bf relevant\_code & \bf method\_body & \bf signature \\
\midrule
\Detailed & 100.00 & 100.00 & 100.00 & 21.93 & 40.24 \\ 
\Sketchy & 100.00 & 100.00 & 100.00 & 29.35 & 45.07 \\ 
\Detailed$_{NoRelatives}$ & 100.00 & 100.00 & 100.00 & 25.32 & 40.51 \\ 
\Sketchy$_{NoRelatives}$ & 100.00 & 100.00 & 100.00 & 43.66 & 46.48 \\ 
\bottomrule
\end{tabular}
\caption{Percentage of tasks across which, the corresponding tool has been used at least once. Note that tasks that pass in the first iteration, without needing to call any tools are not considered here. The tools which were mentioned in the few-shot example dominate the table, being called at least once in each task. Many of these calls end up being superfluous.}
\label{tab:rq4_tool_presence}
\end{table}

%% file: sections/research_questions/rq6.tex
\subsubsection{RQ6 - Number of iterations}

\begin{table}[h]
    \centering
    \begin{tabular}{lccccccc}
        \toprule    
        \bf Language & \bf Category & \textbf 1 & \bf 2& \bf 3 & \bf 4 & \bf 5 \\
        \midrule
        \multirow{2}{*}{\Java} & \Detailed & 13.85 & 54.74 & 70.51 & 75.26 & 77.56 \\ 
         & \Sketchy & 9.49 & 48.08 & 58.59 & 66.03 & 70.64 \\ 
        \multirow{2}{*}{\CSharp} & \Detailed & 0.28 & 14.69 & 21.19 & 27.40 & 33.05 \\ 
         & \Sketchy & 0.00 & 6.21 & 15.82 & 21.19 & 23.73 \\
        \multirow{2}{*}{\Python} & \Detailed & 17.53 & 23.69 & 27.32 & 29.04 & 29.38 \\
         & \Sketchy & 11.17 & 15.29 & 17.53 & 18.56  & 19.24 \\
    \bottomrule     
    \end{tabular}
    \caption{Pass@1 performance across the 5 iterations of \OurMethod. While the performance improves across iterations, the increase in performance at each iteration diminishes.}
    \label{tab:iteration_analysis}
    \label{tab:performance}
\end{table}

To examine the effect of the number of iterations on the performance, we measure the performance at each iteration of \OurMethod in Table \ref{tab:iteration_analysis}. While, by definition, the performance is strictly non-decreasing with the number of iterations, the performance delta decreases between iterations. Thus, there exists a trade-off between the computational resources required for LLM inference and the gain in performance. While, in our case we terminate the generation at 5 iterations, for other datasets, depending on the complexity of the class, more iterations may be required.

%% file: sections/research_questions/rq7.tex
\subsubsection{RQ7 - Multiple generations}

To test the effect of sampling multiple generations, we also report the performance numbers over multiple generations on the java dataset. Specifically, for $n=6$ generations, we calculate the Pass@1, 2, 3 scores. Tables \ref{tab:p123_detailed} and \ref{tab:p123_sketchy} shows that the trends remain, and \OurMethod~outperforms the baselines significantly under all configurations.

{
\tabcolsep=0.15cm
\begin{table}[h]
\begin{center}
\begin{tabular}{lcccccccccccc}
\toprule
\bf Method & \multicolumn{3}{c}{\bf \Java} & \multicolumn{3}{c}{\bf \CSharp} & \multicolumn{3}{c}{\bf \Python} \\
\cmidrule(lr){2-4}
\cmidrule(lr){5-7}
\cmidrule(lr){8-10}
& \bf P@1 &\bf P@2 & \bf P@3 & \bf P@1 &\bf P@2 & \bf P@3 & \bf P@1 &\bf P@2 & \bf P@3  \\
\midrule
\BasicPrompting & 2.18 & 2.29 & 2.30 & 0.00 & 0.00 & 0.00 & 1.2 & 1.37 & 1.55 \\ 
\Reflexion & 18.59 & 22.67 & 25.02 & 2.54 & 3.53 & 3.98 & 8.42 & 12.16 & 14.85 \\ 
\NaiveRAG & 14.23 & 14.72 & 14.90 & 1.69 & 1.69 & 1.69 & 17.01 & 18.83 & 19.38 \\ 
\Repocoder & 54.74 & 56.05 & 56.65 & 1.98 & 2.21 & 2.41 & 23.54 & 26.25 & 27.27 \\ 
\OurMethod & 77.56 & 81.22 & 82.77 & 33.05 & 37.24 & 39.57 & 29.38 & 33.33 & 35.21 \\ 
\bottomrule
\end{tabular}
\end{center}
\caption{Pass @ 1,2,3 scores for the \Detailed~version of \datasetname~setting the number of generations n=6}
\label{tab:p123_detailed}
\end{table}
}

{
\tabcolsep=0.15cm
\begin{table}[h]
\begin{center}
\begin{tabular}{lcccccccccccc}
\toprule
\bf Method & \multicolumn{3}{c}{\bf \Java} & \multicolumn{3}{c}{\bf \CSharp} & \multicolumn{3}{c}{\bf \Python} \\
\cmidrule(lr){2-4}
\cmidrule(lr){5-7}
\cmidrule(lr){8-10}
& \bf P@1 &\bf P@2 & \bf P@3 & \bf P@1 &\bf P@2 & \bf P@3 & \bf P@1 &\bf P@2 & \bf P@3  \\
\midrule
\BasicPrompting & 0.90 & 1.00 & 1.09 & 0.00 & 0.00 & 0.00 & 0.69 & 1.17 & 1.49 \\ 
\Reflexion & 11.03 & 14.19 & 15.97 & 3.39 & 3.86 & 4.10 & 5.67 & 8.11 & 9.28 \\ 
\NaiveRAG & 9.23 & 9.83 & 10.13 & 0.00 & 0.00 & 0.00 & 11.0 & 12.50 & 12.98 \\ 
\Repocoder & 42.05 & 43.80 & 44.65 & 0.28 & 0.52 & 0.71 & 18.04 & 19.73 & 20.62 \\ 
\OurMethod & 70.64 & 74.77 & 76.58 & 23.73 & 28.15 & 30.32 & 19.24 & 20.96 & 21.6 \\ 
\bottomrule
\end{tabular}
\end{center}
\caption{Pass @ 1,2,3 scores for the \Sketchy~version of \datasetname~setting the number of generations n=6}
\label{tab:p123_sketchy}
\end{table}
}

%% file: sections/research_questions/rq8.tex
\subsubsection{RQ8 - Does performance depend on whether the LLM has seen the repository before?}
The Python dataset has tasks from a repository called Litestar which was created on Github after the training-date cutoff for the GPT models used for evaluation. We notice that the performance trends (Table \ref{tab:rq7}) across different methods remains the same. The smaller size of the classes from Litestar as compared to the other Python repositories might be one of the reasons why all the methods perform better on tasks from Litestar.

\begin{table}[h]

\begin{center}
\begin{tabular}{lccccccc}
\toprule
\bf Method & \multicolumn{2}{c}{\bf \Python - All} & \multicolumn{2}{c}{\bf \Python - Litestar}\\
\cmidrule(lr){2-3}
\cmidrule(lr){4-5}
& \bf P@1 &\bf TR & \bf P@1 &\bf TR  \\
\midrule

\BasicPrompting & 1.2 & 3.1 & 2.4  & 3.42  \\ 
\Reflexion  & 8.42 & 12.93  &  14.23 & 17.2  \\ 
\NaiveRAG &  17.01  & 17.73  & 30.89 & 31.7 \\ 
\Repocoder &  23.54 & 26.83 & 37.8 & 40.65  \\ 
\OurMethod & 29.38 & 35.69 & 44.72 &  52.98 \\

\bottomrule
\end{tabular}
\end{center}
\caption{Performance scores on the Python dataset for (1) all the tasks, (2) for only the tasks from Litestar}
\label{tab:rq7}

\end{table}

%% file: sections/research_questions/rq9.tex
\subsubsection{RQ9 - Other models}
To demonstrate the power of \OurMethod~on other models, we conduct experiments on the \Java~dataset using Phi3-mini-128k-instruct, Phi3-medium-128k-instruct, GPT-3.5-turbo-instruct and LLama3-70b-instruct as can be seen in Table \ref{tab:rq9_detailed} and \ref{tab:rq9_sketchy}. \OurMethod~outperforming the baselines is visible for other models as well. Llama-3-70b-instruct demonstrates strong reasoning and tool usage abilities followed by GPT-3.5-turbo-instruct. The Phi3 models though much smaller, are still able to use tools and improve their performance over the \Repocoder~baseline.

\begin{table}[h]
\small
\begin{center}
\begin{tabular}{lccccccccccccccccc}
\toprule
\bf Method & \multicolumn{3}{c}{\bf Phi3-mini-128k} & \multicolumn{3}{c}{\bf  Phi3-medium-128k} & \multicolumn{3}{c}{\bf  GPT-3.5-turbo} & \multicolumn{3}{c}{\bf  Llama3-70b} \\
\cmidrule(lr){2-4}
\cmidrule(lr){5-7}
\cmidrule(lr){8-10}
\cmidrule(lr){11-13}
& \bf P@1 &\bf TR & \bf CR & \bf P@1 &\bf TR & \bf CR & \bf P@1 &\bf TR & \bf CR & \bf P@1 &\bf TR & \bf CR \\
\midrule
\BasicPrompting & 1.53 & 1.53 & 1.53 & 2.3 & 2.3 & 2.3 & 1.54 & 1.54 & 2.31 & 1.55 & 1.55 & 2.33 \\ 
\Reflexion & 3.82 & 4.77 & 5.34 & 3.45 & 3.45 & 3.45 & 3.85 & 5.04 & 5.38 & 6.98 & 9.24 & 10.08 \\ 
\NaiveRAG & 6.11 & 6.83 & 6.87 & 12.64 & 12.64 & 12.64 & 11.54 & 12.15 & 14.62 & 7.75 & 7.75 & 8.53 \\ 
\Repocoder & 10.69 & 11.41 & 11.45 & 44.83 & 46.8 & 47.13 & 40.77 & 43.38 & 46.92 & 31.01 & 32.27 & 33.33 \\ 
\OurMethod & 19.08 & 20.77 & 22.14 & 52.87 & 56.55 & 60.92 & 54.62 & 63.22 & 70.77 & 73.64 & 79.98 & 86.05 \\ 
\bottomrule
\end{tabular}
\end{center}
\caption{Performance numbers expressed in percentage, for the baselines and \OurMethod~on the \Detailed~version of the \Java~dataset. The trend observed on GPT-4 holds for other models as well. It is important to note that minor changes were made to the prompts to cater the requirements of individual models.}
\label{tab:rq9_detailed}
\end{table}

\begin{table}[h]
\small
\begin{center}
\begin{tabular}{lccccccccccccccccc}
\toprule
\bf Method & \multicolumn{3}{c}{\bf Phi3-mini-128k} & \multicolumn{3}{c}{\bf  Phi3-medium-128k} & \multicolumn{3}{c}{\bf  GPT-3.5-turbo} & \multicolumn{3}{c}{\bf  Llama3-70b} \\
\cmidrule(lr){2-4}
\cmidrule(lr){5-7}
\cmidrule(lr){8-10}
\cmidrule(lr){11-13}
& \bf P@1 &\bf TR & \bf CR & \bf P@1 &\bf TR & \bf CR & \bf P@1 &\bf TR & \bf CR & \bf P@1 &\bf TR & \bf CR \\
\midrule
\BasicPrompting & 0.76 & 1.07 & 1.53 & 1.11 & 2.11 & 3.33 & 1.54 & 1.54 & 2.31 & 0.78 & 1.47 & 3.88 \\ 
\Reflexion & 1.53 & 2.21 & 3.05 & 1.11 & 2.11 & 3.33 & 2.31 & 3.04 & 5.38 & 8.53 & 11.34 & 15.5 \\ 
\NaiveRAG & 2.29 & 2.29 & 3.05 & 8.89 & 10.24 & 11.11 & 8.46 & 8.46 & 10 & 4.65 & 5.22 & 6.2 \\ 
\Repocoder & 6.11 & 7.67 & 9.92 & 32.22 & 40.09 & 45.56 & 34.62 & 39.17 & 44.62 & 26.36 & 29.49 & 33.33 \\ 
\OurMethod & 13.74 & 16.8 & 22.14 & 45 & 50.69 & 57.22 & 48.46 & 54.72 & 64.62 & 63.57 & 71.64 & 85.27 \\ 

\bottomrule
\end{tabular}
\end{center}
\caption{Performance numbers expressed in percentage, for the baselines and \OurMethod~on the \Sketchy~version of the \Java~dataset. The trend observed on GPT-4 holds for other models as well. It is important to note that minor changes were made to the prompts to cater the requirements of individual models.}
\label{tab:rq9_sketchy}
\end{table}

%% file: sections/appendix/dataset_generation.tex
\section{\datasetname}

\input{tables_used/dataset_stats}

\subsection{Benchmark Construction}

\subsubsection{Repository Selection}
\label{sec:repository-selection}
Our dataset comprises repositories from two distinct categories: \textbf{(Type 1)} well-established repositories such as `scikit-learn`, `requests`, `pydicom`, which have been present on GitHub since before September 2021; and \textbf{(Type 2)} repositories that were created on GitHub after the cutoff date of the language models (LMs) we are using, i.e., September 2021, ensuring that the LMs have not been exposed to these repositories during their training or fine-tuning phases. We detail the language-specific selection process below:
\begin{itemize}
    \item \textbf{\Java}: All \Java ~repositories included in our study are from (Type 2). To construct the Java dataset, we utilized the existing dataset compiled by MGD \citealp{bench:mgd}.
    \item \textbf{\CSharp}: All \CSharp repositories included in our study are from (Type 2). To construct the Java dataset, we utilized the \textit{StabilityMatrix} repository.    
    \item \textbf{\Python}: For (Type 1) repositories, we adapted the established SWEBench \citep{swebench} benchmark. To mitigate the risk of dataset contamination, we paraphrased the symbols in these repositories as described in Section \ref{sec:dataset-paraphrasing}. To assemble a pool of (Type 2) repositories, we identified the most starred Python repositories on GitHub created after the LMs' cutoff date. We then excluded repositories without any mention of `pytest` in their files, assuming the absence of test cases. Many of the remaining repositories were associated with LMs and appeared to require an `OPENAPI\_KEY` to execute tests. Consequently, we excluded repositories containing the keywords `OPEN\_API`, `LLM`, and `GPT`. From the remaining candidates, we selected the top three repositories that did not seem to be related to LMs based on their title or description. This process yielded three repositories: `dosisod/refurb`, `pyscript/pyscript`, and `litestar-org/litestar`.
\end{itemize}

\subsubsection{Task instance construction from each repository}
\label{sec:class_shortlist}
For each repository identified in the previous section, we first ensure that we can successfully build the repository (for \Java~and \CSharp) and that all the necessary environment installations are in place (for \Python). We provide the necessary scripts to install such conda environments wherever applicable. Once these prerequisites are met, we refer to the current state of the repository as `R` and begin the process of shortlisting candidate classes. An ideal class for inclusion in our dataset should exhibit two key properties: (1) it utilizes context from the repository, and (2) its correctness can be verified through test cases. We define these properties in more detail as follows:

\begin{itemize}
    \item \textbf{Uses Repository-Level Context}: For a given class `C`, there can be four types of references in its body:
    \begin{enumerate}
        \item References defined in external libraries outside the repository.
        \item References to other members within the class `C` itself (e.g., method `M1` of class `C` calling method `M2` of the same class `C`).
        \item References to entities defined in the same file but outside the body of class `C`.
        \item References to entities defined elsewhere within the repository.
    \end{enumerate}
    For our study, we categorize references of type (3) and (4) as \textbf{EXTERNAL REFERENCES}.

    \item \textbf{Covered Under Test Cases}: For repository $R$, let $P$ denote the set of test cases that pass in the current state. Given a test case $T$ and a class $C$ from our benchmark tasks, we define:

\end{itemize}
\begin{align}
\begin{split}
    \text{DirectCoverage}(T) = &\text{ the set of classes/functions} \\
    &\text{directly invoked in the body of } T,
\end{split} \label{coverage:m1} \\
\begin{split}
    \text{IndirectCoverage}(T) = &\text{ the set of classes/functions not} \\
    &\text{directly invoked in the body}\\
    &\text{of T, but if left unimplemented, }\\
    &\text{T would fail}
\end{split} \label{coverage:m2}
\end{align}
    We then determine the number of unique test cases that directly cover class $C$ or any of its members.

To confirm that each class possesses both of the aforementioned properties, we apply the following filtering criteria across different languages:
\begin{itemize}
    \item \textbf{\Java}: We require that at least two-thirds of the methods in a class are referenced in the combined bodies of all corresponding test cases to ensure adequate code coverage. Additionally, the class must contain at least one external reference.
    \item \textbf{\Python}: We require at least two methods that are (a) directly covered by some test case and (b) make an external reference.
    \item \textbf{\CSharp}: We require that at least 30\% of the lines in the file are covered by the test cases and the class contains at least one external reference.    
\end{itemize}

\subsubsection{Dataset Paraphrasing}
\label{sec:dataset-paraphrasing}
For each repository `R`, we begin by compiling a list of all identifiers that appear in at least one class within our benchmark. To paraphrase these identifiers, we apply a case-flipping technique to their original names. For example, the identifier `encode` would be transformed to `eNcoDe`. This transformation is applied consistently across all Python and Cython files in the repository. \citep{mxeval} has shown the LMs are not robust to errors when prone to natural language descriptions with randomly flipped characters. 

To avoid inadvertently altering identifiers from external libraries, we exclude certain common identifiers from this process. For instance, we would not modify the identifier `items` to prevent the expression `my\_dict.items()` from being incorrectly changed to `my\_dict.ItEms()`.

We verify the success of our paraphrasing by ensuring that the majority of test cases that passed prior to the paraphrasing continue to pass afterwards. This approach to paraphrasing serves a dual purpose: (1) it preserves the semantic meaning of the original identifiers, and (2) it prevents the LLM from relying on rote memorization of its training data to complete tasks in our benchmark. A sample instance of paraphrased code can be found in Diff \ref{diffs:example2}.

\input{tables_used/metadata_contents}
\subsection{Task components}
\textbf{Repository setup}: At the start of each task, the repository is reset to its original state, ensuring all components are aligned with the ground truth. The class targeted by the task, along with its associated imports, is then removed from the repository.

\textbf{Model input}: The model receives the NL description of the class and is tasked with generating the complete class body, including any necessary import statements.

\textbf{Testcase Feedback}: We identify test cases that pass in the repository's ground truth state and also reference the class or its members within their test functions. An incorrect class implementation could lead to failures in these tests. Let's call this set \expectedpass{}.`The model's output is assessed against this specific set of test cases.

\textbf{Evaluation metrics}: To evaluate the model's generated code, we insert it into the repository at the location of the original class implementation and run the relevant test cases (ie \expectedpass{}). Our evaluation metrics include: To score the generation, we use the following metrics:
\begin{itemize}
    \item \textbf{Testcase pass rate (TR)}: This is the fraction of test cases from \expectedpass{} which pass when the model's code is introduced into the repository.
    \item \textbf{Compilation Rate (CR)}: This is a binary value between 1 and 0 depending on whether the repository was able to be built after the model's code was introduced in the repository. (Applicable only to Java dataset)
\end{itemize}

%% file: tables_used/dataset_stats.tex
\begin{table}[!ht]
    \centering

    \begin{tabular}{lcccc}
    \toprule
        & \bf \Java & \bf \Python &\bf \CSharp\\
        \midrule
        Num. of tasks & 130 & 97 & 60 \\        
        Length of \Detailed~NL description & 1475.98 / 286.89 & 3245.23 / 771.77 & 3322.15 / 657.58 \\
        Length of \Sketchy~NL description & 1481.69 / 269.81 & 2633.20 / 607.64 & 1079.13 / 211.30\\
        Length of classes & 2080 / 452.69 & 4663.76 / 1070.49 & 4283.00 / 842.62 \\ 
        Num. of TCs directly covering the classes & 5.48 & 42.94 & 2.27 \\
        Num. of unique Ext. Refs& 3.51 & 7.06 & -\\
        Num. of funcs in the class & 3.1 & 9.29 & 4.15 \\
        Num. of funcs covered in at least one TC & 2.85 & 4.84 & - \\ 
        Num. of funcs making at least one Ext. Refs & 2.28 & 4.84 & -\\ 
    \bottomrule
    \end{tabular}
    \caption{Dataset High level Statistics. Each row represents an average over all the tasks in the dataset. The cells with $/$ represent the $<$number of characters$>$ / $<$number of tokens using gpt-3.5 tokenizer$>$. TC = Test Cases, funcs = functions, Ext. Refs = References from other files in the repository. We measure test coverage on \CSharp~using a utility called Coverlet, whose output contains a file-specific percentage measure of coverage on each file for each test case, therefore function-level measures are hard to calculate.  }
    \label{tab:dataset_high_level_statistics}
\end{table}

%% file: tables_used/metadata_contents.tex
\renewcommand{\arraystretch}{1.1}
\begin{table*}[htp]
    \centering
    
    \begin{tabular}{l|l|c|c}
        \hline
        \textbf{Level} & \textbf{Field} & \Descriptive & \Sketchy \\ \hline
        \multirow{7}{*}{Class Level Info} & 
        Class Name and file path & \greencheck & \greencheck \\
        & Import statements & \redcross & \redcross \\
        & Member variables (with initializations) & \greencheck & \greencheck \\
        & Class signature & \greencheck & \greencheck \\
        & Decorators & \greencheck & \greencheck \\ 
        & Parent Class names & \greencheck & \greencheck \\ \hline
        \multirow{4}{*}{Method Level Info} & 
        Method Signature & \greencheck & \greencheck \\
        & Method decorators & \greencheck & \greencheck \\
        & Method Docstrings & \greencheck & \greencheck \\
        & Method Body & \greencheck & \redcross \\
        \hline
    \end{tabular}
    \caption{Components of the metadata}    
    \label{table:metadata_components}
\end{table*}

%% file: sections/appendix/tool_descriptions.tex
\section{Tools}

Broadly speaking, the tools utilized in \OurMethod~are classified as either dependent or independent, depending on whether their invocation requires reasoning. Independent tools do not necessitate reasoning about the current state of the \OurMethod~loop. These tools are automatically called during the initial generation phase. Our independent toolset contains a single tool get\_related\_snippets. On the other hand, tools requiring reasoning over the current state of the RepoReflexion loop are classified as dependent tools. Our dependent toolset contains get\_imports, get\_class\_info, get\_signature, get\_method\_body and get\_relevant\_code/.

\subsection{Independent Tools}
Independent tools do not necessitate reasoning about the current state of the \OurMethod~loop. These tools are automatically called during the initial generation phase.
\begin{itemize}
\item \textbf{get\_related\_snippets}: This tool addresses the common scenario of multiple similar classes within a repository. It segments the repository into snippets and returns the top 5 snippets based on cosine similarity with the class description. Since repositories often contain near-identical classes, the LLM agent benefits from examining these implementations.
\end{itemize}

\subsection{Dependent Tools}
Tools requiring reasoning over the current state of the RepoReflexion loop are classified as dependent tools. \begin{itemize}
    \item \textbf{get\_imports} (Parameters: Empty): This tool suggests imports all the undefined symbols in the current generation. It scans the repository for potential source files defining the symbol and recommends import statements. If multiple sources are possible, it outputs all options for the LLM agent to choose from. The import tool can be helpful to resolve "symbol not found errors".
    
    \item \textbf{get\_class\_info} (Parameter 1= Class Name): This tool locates the class definition in the repository and gathers information about its members, including inherited members. The tool provides detailed information about each member, such as parameters, return type, access specifier, and whether the member is static or abstract. This tool can help when a method/variable in a class returns a "symbol not found" error In cases of multiple classes with the same name, the tool lists information for each. To manage prompt length, the tool ranks members based on cosine similarity with the thought produced just before invocation, displaying the top k results, where k is set to 10. In case multiple classes with the same name exist, it shows the info for each of them. In case the LLM passes multiple classes, it shows the info for each of them.
    \item \textbf{get\_signature} (Parameter 1= Class Name, Parameter 2= Method Name): This tool returns the signature of the requested method. In case multiple methods with the same name exist in the same class (overloading), it displays the signatures of all of them. This could help when the number or parameters or the types of the parameters were hallucinated, leading to incorrect method calls. 
    \item \textbf{get\_method\_body} (Parameter 1= Class Name, Parameter 2= Method Name): This tool checks the source code and returns the method definition of the requested method. In case it is too large, it truncates the output. In case multiple methods with the same name are available, it shows the definition for each of them. In case the definition is unavailable (due to the method being in an external library), it shows the signature instead. This tool can be invoked to address the situation where the exact implementational logic of a method is required to fix an error.
    \item \textbf{get\_relevant\_code} (Parameter 1= Query String): While all the tools till now can help gather more information about symbols that the LLM already has knowledge about, a large part of writing code involves dependencies that the coder is unaware of. For instance, there may be methods in utility classes that can be re-used instead of writing the logic from scratch. To aid in the search for relevant clues across the repository, this tool allows the LLM to make specific queries that retrieve code structures using embedding similarity scores. The tool considers three types of code-pieces, classes, independent methods (not present in a class) and snippets. For classes, the methods are stripped of their bodies before the encoding, for independent methods (not present in a class), the body of the method is used, and for snippets, the snippets are used directly to generate the embeddings. The tool returns the top 3 structures on the basis of the cosine similarity score.
\end{itemize}

\subsection{Implementation details}
For Python, all the tools were implemented using a combination of Jedi and Tree-Sitter. For Java, the EclipseJDTLS Language-server was used.

%% file: motivating_examples/example_1/run_1.tex
\textbf{Model}: GPT-4 \\
\textbf{Task Instance:} A class generation task from the Java repository called \textit{gchq/sleeper}\\
\\
\textbf{Discussion:} For the task, both the \hyperlink{eg:detailedprompt}{detailed} and \hyperlink{eg:sketchyprompt}{sketchy prompts} can be found below. The highlighted portion of the detailed prompt has a mention of keywords such as \textit{rowKeyNames} and \textit{sortKeyNames}, whereas the same portion of the sketchy prompt does not mention these names specifically. Due to the keywords, one of the snippets retrieved by RepoCoder for the detailed prompt is \hyperlink{eg:snippetretrieved}{here}, which mentions the words \textit{rowKeyFieldNames} and \textit{sortKeyFieldNames}. This allows both RepoCoder and RRR to generate \hyperlink{eg:detailedcodegen}{correct code} in the first attempt by guessing the correct names of the relevant functions \textit{getRowKeyFieldNames} and \textit{getSortKeyFieldNames}.\\

However, in the sketchy setting, due to the less verbose nature of the description, the snippets retrieved are not helpful. As a result, the \hyperlink{eg:stage0codegen}{code generated} by RepoCoder fails on all 5 attempts. However, RRR is able to recover from this error in the following manner:
\begin{itemize}
    \item It reasons about the \hyperlink{eg:stage0feedback}{compiler feedback} and realizes that it has hallucinated method names \textit{getRowKey} and \textit{getSortKey} for the \textit{Schema} class. \item It decides to \hyperlink{eg:stage0toolinvoke}{invoke tool} \textit{get\_class\_info(Schema)} to get the list of all valid methods of the \textit{Schema} class.
    \item In the \hyperlink{eg:stage0tooloutput}{output returned by the tools}, it comes to know about the occurrence of correct methods \textit{getRowKeyFieldNames} and \textit{getSortKeyFieldNames}. It \hyperlink{eg:stage0reasontooloutput}{reasons about how it's mistake can be corrected} by using the correct methods and makes another attempt to generate code.
    \item In it's \hyperlink{eg:stage1codegen}{2nd code-generation attempt}, it receives \hypertarget{eg:stage1feedback}{compiler feedback} that it was incorrectly trying to cast the output of \textit{Record.get} method to a \textit{Comparable}. 
    \item It \hyperlink{eg:stage1toolinvoke}{invokes} the tool \textit{get\_class\_info(Record)} whose \hyperlink{eg:stage1tooloutput}{output} enables it to \hyperlink{eg:stage1reasontooloutput}{reason} that the return type of \textit{Record.get} is an \textit{Object} which cannot be cast to \textit{Comparable} type.
    \item Using the tool outputs and reflection feedback, the LLM is able to corect it's mistake and \hyperlink{eg:stage2codegen}{generate code} which passes all testcases.
\end{itemize}

\noindent\rule{\textwidth}{0.4pt}
\textit{\textbf{Detailed Description}} \\
\hypertarget{eg:detailedprompt}{}
\input{motivating_examples/example_1/00_detailed_description} 
% \midrule
\noindent\rule{\textwidth}{0.4pt}
\textit{\textbf{Sketchy Description}} \\
\hypertarget{eg:sketchyprompt}{}
\input{motivating_examples/example_1/01_sketchy_description} 
\noindent\rule{\textwidth}{0.4pt}
\textbf{Ground Truth}
\input{motivating_examples/example_1/02_ground_truth_code}
%%%%%%%%%%%%%%%%%%%%%%%%%%%%%%
\noindent\rule{\textwidth}{0.4pt}
\hypertarget{eg:snippetretrieved}{}
\textbf{[Detailed Setting] Snippet retrieved using embedding based tools}\\
\input{motivating_examples/example_1/04_snippet_retrieved_which_helped} 
%%%%%%%%%%%%%%%%%%%%%%%%%%%%%%%%%%%
\noindent\rule{\textwidth}{0.4pt}
\hypertarget{eg:detailedcodegen}{}
\textbf{[Detailed Setting] Stage 0: Code generated by RepoCoder (and RRR) in their 1st attempts which is Correct}\\
\input{motivating_examples/example_1/03_detailed_code_generated} 
%%%%%%%%%%%%%%%%%%%%%%%%%%%%%%%
\noindent\rule{\textwidth}{0.4pt}
\hypertarget{eg:stage0codegen}{}
\textbf{[Sketchy Setting] Stage 0: Code generated by RepoCoder (and RRR) in their first attempt which is incorrect}\\
\input{motivating_examples/example_1/05_sketchy_generation_by_repocoder} 
%%%%%%%%%%%%%%%%%%%%%%%%%%%%%%%
\noindent\rule{\textwidth}{0.4pt}
\textbf{[Sketchy Setting] Feedback received by RRR on it's incorrect code generation attempt in stage 0}\\
\hypertarget{eg:stage0feedback}{}
\input{motivating_examples/example_1/06_repotools_feedback_received} 
%%%%%%%%%%%%%%%%%%%%%%%%%%%%%%%
\noindent\rule{\textwidth}{0.4pt}
\textbf{[Sketchy Setting] Stage 1: LLM reflects on the compiler feedback logs and reasons what tools to invoke}\\
\hypertarget{eg:stage0toolinvoke}{}
\input{motivating_examples/example_1/07_tools_called_for_error} 
%%%%%%%%%%%%%%%%%%%%%%%%%%%%%%%
\noindent\rule{\textwidth}{0.4pt}
\hypertarget{eg:stage0tooloutput}{}
\textbf{[Sketchy Setting] The outputs fetched by the tools invoked by the LLM in stage 1}\\
\input{motivating_examples/example_1/08_tool_output} 
%%%%%%%%%%%%%%%%%%%%%%%%%%%%%%%
\noindent\rule{\textwidth}{0.4pt}
\hypertarget{eg:stage0reasontooloutput}{}
\textbf{[Sketchy Setting]  Stage 2: the LLM reasons how the results of the invoked tools can be used to correct it's mistake}\\
\input{motivating_examples/example_1/09_reflection_on_tool_output}
%%%%%%%%%%%%%%%%%%%%%%%%%%%%%%%
\noindent\rule{\textwidth}{0.4pt}
\textbf{[Sketchy Setting] Stage 3: 2nd code-generation attempt by the LLM with the tool outputs and compiler feedback of the earlier attempt in context}
\hypertarget{eg:stage1codegen}{}
\input{motivating_examples/example_1/10_codegen_attempt_2} 
%%%%%%%%%%%%%%%%%%%%%%%%%%%%%%%
\noindent\rule{\textwidth}{0.4pt}
\textbf{[Sketchy Setting] Feedback received by RRR on it's incorrect code generation attempt in stage 3}\\
\hypertarget{eg:stage1feedback}{}
\input{motivating_examples/example_1/11_feedback_on_attempt_2} 
%%%%%%%%%%%%%%%%%%%%%%%%%%%%%%%
\noindent\rule{\textwidth}{0.4pt}
\textbf{[Sketchy Setting] Stage 4: LLM reflects on the compiler feedback logs and reasons what tools to invoke}\\
\hypertarget{eg:stage1toolinvoke}{}
\input{motivating_examples/example_1/12_tools_invoked_attempt_2} 
%%%%%%%%%%%%%%%%%%%%%%%%%%%%%%%
\noindent\rule{\textwidth}{0.4pt}
\textbf{[Sketchy Setting] The outputs fetched by the tools invoked by the LLM in stage 4}\\
\hypertarget{eg:stage1tooloutput}{}
\input{motivating_examples/example_1/13_reason_tool_output} 
%%%%%%%%%%%%%%%%%%%%%%%%%%%%%%%
\noindent\rule{\textwidth}{0.4pt}
\textbf{[Sketchy Setting] Stage 5: the LLM reasons how the results of the invoked tools can be used to correct it's mistake}\\
\hypertarget{eg:stage1reasontooloutput}{}
\input{motivating_examples/example_1/14_reflection_on_tool_output_attempt_2} 
% %%%%%%%%%%%%%%%%%%%%%%%%%%%%%%%
\noindent\rule{\textwidth}{0.4pt}
\textbf{[Sketchy Setting] Stage 6: 2nd code-generation attempt by the LLM with the tool outputs and compiler feedback of the earlier attempt in context [correct attempt => terminate RRR loop]}\\
\hypertarget{eg:stage2codegen}{}
\input{motivating_examples/example_1/15_codegen_attempt_3} 
%%%%%%%%%%%%%%%%%%%%%%%%%%%%%%%
%     \bottomrule
% \end{tabular}
% \end{table}

%% file: motivating_examples/example_1/00_detailed_description.tex
\textbf{File:} sleeper/core/record/RecordComparator.java\\
\textbf{Description:} The public class \pythoninline{RecordComparator} implements the Comparator interface for objects of the Record class. This class is designed to provide comparison logic for Record objects based on a schema.

\colorbox{yellow}{The class has four private instance variables: \textit{rowKeyNames}, \textit{sortKeyNames}, \textit{rowKeyComparator}, and \textit{sortKeyComparator}.} \colorbox{yellow}{The \textit{rowKeyNames} and \textit{sortKeyNames} are lists of strings that store the names of the row keys and sort keys respectively.} The \textit{rowKeyComparator} and \textit{sortKeyComparator} are instances of the KeyComparator class, which are used to compare the row keys and sort keys.

The class has a public constructor that takes a \pythoninline{Schema} object as an argument. This constructor initializes the \textit{rowKeyNames} and \textit{sortKeyNames} variables with the row key field names and sort key field names from the schema respectively. It also initializes the rowKeyComparator and sortKeyComparator variables with new \pythoninline{KeyComparator} objects, using the row key types and sort key types from the schema respectively.

The class overrides the compare method from the \pythoninline{Comparator} interface. This method takes two Record objects as arguments and compares them first by their row keys and then by their sort keys. It retrieves the values of the row keys and sort keys from the records, creates Key objects from these values, and then compares these keys using the \textit{rowKeyComparator} and \textit{sortKeyComparator}. If the row keys are not equal, it returns the result of the row key comparison. If the row keys are equal, it returns the result of the sort key comparison. This method is used to determine the order of Record objects based on their row keys and sort keys.

%% file: motivating_examples/example_1/01_sketchy_description.tex
\textbf{File:} sleeper/core/record/RecordComparator.java\\
\textbf{Description:} The public class \pythoninline{RecordComparator}, which implements the \pythoninline{Comparator} interface for Record objects, is designed to provide a custom comparison logic that is based on a given schema. This class is particularly useful when sorting or ordering collections of Record objects according to specific criteria defined by the schema. \colorbox{yellow}{The class contains private final variables that hold the names of the row keys and sort keys, as well as comparators for these} \colorbox{yellow}{keys, which are used in the comparison process.}

The constructor of the RecordComparator class takes a Schema object as an argument and uses this schema to initialize the comparators for both the row keys and the sort keys. These comparators are then used to define the ordering of the Record objects.

The \pythoninline{compare} method is a crucial part of the class, as it overrides the compare method from the Comparator interface. It takes two Record objects as parameters and compares them by first looking at their row keys and then at their sort keys. The comparison is performed in a way that respects the ordering defined by the schema, and the method returns an integer value that indicates the relative ordering of the two records. A negative value means the first record is less than the second, zero means they are equal, and a positive value means the first record is greater than the second. This method enables the sorting of Record objects in collections such as lists or trees according to the specified schema.

%% file: motivating_examples/example_1/02_ground_truth_code.tex
\begin{lstlisting}[language=Java]
import sleeper.core.key.Key;
import sleeper.core.schema.Schema;

import java.util.Comparator;
import java.util.List;

/**
 * Compares Records by row keys then sort keys.
 */
public class RecordComparator implements Comparator<Record> {
    private final List<String> rowKeyNames;
    private final List<String> sortKeyNames;
    private final KeyComparator rowKeyComparator;
    private final KeyComparator sortKeyComparator;

    public RecordComparator(Schema schema) {
        this.rowKeyNames = schema.getRowKeyFieldNames();
        this.sortKeyNames = schema.getSortKeyFieldNames();
        this.rowKeyComparator = new KeyComparator(schema.getRowKeyTypes());
        this.sortKeyComparator = new KeyComparator(schema.getSortKeyTypes());
    }

    // TODO Optimise by avoiding creating lists of row keys and sort keys, and
    // just do the comparison directly here?
    @Override
    public int compare(Record record1, Record record2) {
        List<Object> record1Key = record1.getValues(rowKeyNames);
        List<Object> record2Key = record2.getValues(rowKeyNames);
        int keyComparison = rowKeyComparator.compare(Key.create(record1Key), Key.create(record2Key));
        if (0 != keyComparison) {
            return keyComparison;
        }
        List<Object> record1SortFields = record1.getValues(sortKeyNames);
        List<Object> record2SortFields = record2.getValues(sortKeyNames);
        return sortKeyComparator.compare(Key.create(record1SortFields), Key.create(record2SortFields));
    }
}
\end{lstlisting}

%% file: motivating_examples/example_1/04_snippet_retrieved_which_helped.tex
\begin{lstlisting}
####SNIPPET 1
####From file src/main/java/sleeper/core/iterator/impl/AdditionIterator.java:
```        @Override
        public void close() throws IOException {
            input.close();
        }
    }

    private static boolean equalRowAndSort(List<String> rowKeyFieldNames,
                                           List<String> sortKeyFieldNames, Record record1, Record record2) {
        List<Object> keys1 = new ArrayList<>();
        List<Object> keys2 = new ArrayList<>();
        for (String rowKey : rowKeyFieldNames) {
            keys1.add(record1.get(rowKey));
            keys2.add(record2.get(rowKey));
        }
        for (String sortKey : sortKeyFieldNames) {
            keys1.add(record1.get(sortKey));
            keys2.add(record2.get(sortKey));
        }
        return Key.create(keys1).equals(Key.create(keys2));
    }
\end{lstlisting}

%% file: motivating_examples/example_1/03_detailed_code_generated.tex
\begin{lstlisting}[language=Java]
import sleeper.core.record.Record;
import sleeper.core.schema.Schema;
import sleeper.core.schema.type.PrimitiveType;
import sleeper.core.key.Key;
import java.util.Comparator;
import java.util.List;

public class RecordComparator implements Comparator<Record> {
    private List<String> rowKeyNames;
    private List<String> sortKeyNames;
    private KeyComparator rowKeyComparator;
    private KeyComparator sortKeyComparator;

    public RecordComparator(Schema schema) {
        this.rowKeyNames = schema.getRowKeyFieldNames();
        this.sortKeyNames = schema.getSortKeyFieldNames();
        this.rowKeyComparator = new KeyComparator(schema.getRowKeyTypes());
        this.sortKeyComparator = new KeyComparator(schema.getSortKeyTypes());
    }

    @Override
    public int compare(Record record1, Record record2) {
        Key rowKey1 = Key.create(record1.getValues(rowKeyNames));
        Key rowKey2 = Key.create(record2.getValues(rowKeyNames));
        int rowKeyComparison = rowKeyComparator.compare(rowKey1, rowKey2);
        if (rowKeyComparison != 0) {
            return rowKeyComparison;
        }

        Key sortKey1 = Key.create(record1.getValues(sortKeyNames));
        Key sortKey2 = Key.create(record2.getValues(sortKeyNames));
        return sortKeyComparator.compare(sortKey1, sortKey2);
    }
}
\end{lstlisting}

%% file: motivating_examples/example_1/05_sketchy_generation_by_repocoder.tex
\begin{lstlisting}[language=Java]
import sleeper.core.record.Record;
import sleeper.core.schema.Schema;
import java.util.Comparator;

public class RecordComparator implements Comparator {
    private final String rowKey;
    private final String sortKey;
    private final Comparator rowKeyComparator;
    private final Comparator sortKeyComparator;

    public RecordComparator(Schema schema) {
        this.rowKey = schema.getRowKey();
        this.sortKey = schema.getSortKey();
        this.rowKeyComparator = Comparator.comparing(Record::getRowKey);
        this.sortKeyComparator = Comparator.comparing(Record::getSortKey);
    }

    @Override
    public int compare(Record record1, Record record2) {
        int rowKeyComparison = rowKeyComparator.compare(record1, record2);
        if (rowKeyComparison != 0) {
            return rowKeyComparison;
        }
        return sortKeyComparator.compare(record1, record2);
    }
}
\end{lstlisting}

%% file: motivating_examples/example_1/06_repotools_feedback_received.tex
\begin{lstlisting}[language=bash]
Failed to execute goal org.apache.maven.plugins:maven-compiler-plugin:3.10.1:compile (default-compile) on project core: Compilation failure: Compilation failure: 
[ERROR] /eval_dir/sleeper/java/core/src/main/java/sleeper/core/record/RecordComparator.java:[29,29] cannot find symbol
[ERROR]   symbol:   method getRowKey()
[ERROR]   location: variable schema of type sleeper.core.schema.Schema
[ERROR] /eval_dir/sleeper/java/core/src/main/java/sleeper/core/record/RecordComparator.java:[30,30] cannot find symbol
[ERROR]   symbol:   method getSortKey()
[ERROR]   location: variable schema of type sleeper.core.schema.Schema
[ERROR] /eval_dir/sleeper/java/core/src/main/java/sleeper/core/record/RecordComparator.java:[31,54] invalid method reference
[ERROR]   cannot find symbol
[ERROR]     symbol:   method getRowKey()
[ERROR]     location: class sleeper.core.record.Record
[ERROR] /eval_dir/sleeper/java/core/src/main/java/sleeper/core/record/RecordComparator.java:[32,55] invalid method reference
[ERROR]   cannot find symbol
[ERROR]     symbol:   method getSortKey()
[ERROR]     location: class sleeper.core.record.Record
[ERROR] /eval_dir/sleeper/java/core/src/main/java/sleeper/core/record/RecordComparator.java:[37,57] incompatible types: sleeper.core.record.Record cannot be converted to java.lang.String
[ERROR] /eval_dir/sleeper/java/core/src/main/java/sleeper/core/record/RecordComparator.java:[41,42] incompatible types: sleeper.core.record.Record cannot be converted to java.lang.String
\end{lstlisting}

%% file: motivating_examples/example_1/07_tools_called_for_error.tex
\begin{lstlisting}
The error messages indicate that the methods getRowKey() and getSortKey() cannot be found in both the Schema and Record classes. To resolve these errors, I need to check the available methods in both classes. Therefore, I will use the get_class_info tool for both classes. Additionally, I need to check if there are any missing imports that could be causing these errors, so I will use the get_imports tool. 

get_class_info('Schema')
get_class_info('Record')
get_imports()
\end{lstlisting}

%% file: motivating_examples/example_1/08_tool_output.tex
\begin{lstlisting}
################
# Suggested imports:
For Schema you can choose between 'javax.xml.validation.Schema' and 'sleeper.core.schema.Schema'
For Comparator you can use 'java.util.Comparator'

 Note that these are the only imports needed based on the current code. No extra imports are required.

################
# For class sleeper.core.record.Record:
Constructor of class sleeper.core.record.Record has signature: 
Record.Record(Record record)
Record.Record(Map map)
Record.Record()

Objects of the class sleeper.core.record.Record have the following instance members: 
-getKeys() : Set   (instance method)
-getValues(java.util.List fieldNames) : List   (instance method)
-remove(java.lang.String fieldName) : void   (instance method)
-get(java.lang.String fieldName) : Object   (instance method)
-put(java.lang.String fieldName, java.lang.Object value) : void   (instance method)
################
# For class javax.xml.validation.Schema:

Objects of the class javax.xml.validation.Schema have the following instance members: 
-newValidator() : Validator   (instance method)
-newValidatorHandler() : ValidatorHandler   (instance method)
################
# For class sleeper.core.schema.Schema:

The class sleeper.core.schema.Schema has the following static members: 
-load(java.nio.file.Path schemaPath) : Schema   (static method)
-sleeper.core.schema.Schema.Builder  (static variable)
-loadFromString(java.lang.String schemaJson) : Schema   (static method)
-builder() : Builder   (static method)

Objects of the class sleeper.core.schema.Schema have the following instance members: 
-getSortKeyFields() : List   (instance method)
-getSortKeyFieldNames() : List   (instance method)
-getRowKeyFields() : List   (instance method)
-getSortKeyTypes() : List   (instance method)
-getRowKeyFieldNames() : List   (instance method)
-getRowKeyTypes() : List   (instance method)
\end{lstlisting}

%% file: motivating_examples/example_1/09_reflection_on_tool_output.tex
\begin{lstlisting}
The previous implementation is incorrect because it's trying to call non-existent methods. The Schema class does not have getRowKey() or getSortKey() methods, instead it has getRowKeyFieldNames() and getSortKeyFieldNames() methods. Similarly, the Record class does not have getRowKey() or getSortKey() methods. The comparators should be initialized based on the correct methods and fields.
\end{lstlisting}

%% file: motivating_examples/example_1/10_codegen_attempt_2.tex
\begin{lstlisting}[language=Java]
import sleeper.core.record.Record;
import sleeper.core.schema.Schema;
import java.util.Comparator;
import java.util.List;

public class RecordComparator implements Comparator {
    private final List rowKeyFieldNames;
    private final List sortKeyFieldNames;

    public RecordComparator(Schema schema) {
        this.rowKeyFieldNames = schema.getRowKeyFieldNames();
        this.sortKeyFieldNames = schema.getSortKeyFieldNames();
    }

    @Override
    public int compare(Record record1, Record record2) {
        for (String key : rowKeyFieldNames) {
            Comparable value1 = (Comparable) record1.get(key);
            Comparable value2 = (Comparable) record2.get(key);
            int comparison = value1.compareTo(value2);
            if (comparison != 0) {
                return comparison;
            }
        }
        for (String key : sortKeyFieldNames) {
            Comparable value1 = (Comparable) record1.get(key);
            Comparable value2 = (Comparable) record2.get(key);
            int comparison = value1.compareTo(value2);
            if (comparison != 0) {
                return comparison;
            }
        }
        return 0;
    }
}
\end{lstlisting}

%% file: motivating_examples/example_1/11_feedback_on_attempt_2.tex
\begin{lstlisting}[language=bash]
[ERROR] Errors: 
[ERROR]   RecordComparatorTest.shouldCompareCorrectlyWithByteArrayRowKeyAndByteArraySortKey:229 >> ClassCast class [B cannot be cast to class java.lang.Comparable ([B and java.lang.Comparable are in module java.base of loader 'bootstrap')
[ERROR]   RecordComparatorTest.shouldCompareCorrectlyWithByteArrayRowKeyAndNoSortKeys:143 >> ClassCast class [B cannot be cast to class java.lang.Comparable ([B and java.lang.Comparable are in module java.base of loader 'bootstrap')
[ERROR]   RecordComparatorTest.shouldCompareCorrectlyWithMultidimensionalByteArrayRowKeyAndMultidimensionalByteArraySortKey:301 >> ClassCast class [B cannot be cast to class java.lang.Comparable ([B and java.lang.Comparable are in module java.base of loader 'bootstrap')
[ERROR] Tests run: 7, Failures: 0, Errors: 3, Skipped: 0
\end{lstlisting}

%% file: motivating_examples/example_1/12_tools_invoked_attempt_2.tex
\begin{lstlisting}
The error is a ClassCastException, which means that the code is trying to cast an object to a class of which it is not an instance. In this case, it seems that the code is trying to cast a byte array to a Comparable, which is not possible. The error occurs in the compare method of the RecordComparator class, when it tries to cast the values of the keys to Comparable. 

To fix this error, I need to understand how the Record class works, especially the get method. I will use the get_class_info tool to get information about the Record class. I also need to understand how the Schema class works, especially the getRowKeyFieldNames and getSortKeyFieldNames methods. I will use the get_class_info tool to get information about the Schema class. 

Finally, I need to understand how the compare method of the RecordComparator class works. I will use the get_method_body tool to get the body of the compare method. 

I will prioritize these actions because they will help me understand the cause of the error and how to fix it. 

get_class_info('Record')
get_class_info('Schema')
get_method_body('RecordComparator', 'compare')
\end{lstlisting}

%% file: motivating_examples/example_1/13_reason_tool_output.tex
\begin{lstlisting}
#########
# There is no compare method in RecordComparator.
#########
# For class javax.xml.validation.Schema:
Objects of the class javax.xml.validation.Schema have the following instance members: 
-newValidator() : Validator   (instance method)
-newValidatorHandler() : ValidatorHandler   (instance method)
#########
# For class sleeper.core.schema.Schema:

The class sleeper.core.schema.Schema has the following static members: 
-builder() : Builder   (static method)
-loadFromString(java.lang.String schemaJson) : Schema   (static method)
-sleeper.core.schema.Schema.Builder  (static variable)
-load(java.nio.file.Path schemaPath) : Schema   (static method)

Objects of the class sleeper.core.schema.Schema have the following instance members: 
-getSortKeyTypes() : List   (instance method)
-getSortKeyFields() : List   (instance method)
-getSortKeyFieldNames() : List   (instance method)
-getRowKeyTypes() : List   (instance method)
-getRowKeyFields() : List   (instance method)
-getRowKeyFieldNames() : List   (instance method)
#########
# The specified class java.lang.Record does not exist.
#########
# For class sleeper.core.record.Record:
Constructor of class sleeper.core.record.Record has signature: 
Record.Record(Record record)
Record.Record(Map map)
Record.Record()

Objects of the class sleeper.core.record.Record have the following instance members: 
-getKeys() : Set   (instance method)
-getValues(java.util.List fieldNames) : List   (instance method)
-get(java.lang.String fieldName) : Object   (instance method)
-remove(java.lang.String fieldName) : void   (instance method)
-put(java.lang.String fieldName, java.lang.Object value) : void   (instance method)
\end{lstlisting}

%% file: motivating_examples/example_1/14_reflection_on_tool_output_attempt_2.tex
\begin{lstlisting}
The error in the previous implementation is due to the fact that byte arrays cannot be cast to Comparable, which is required for the comparison in the compare method. This is causing a ClassCastException during the execution of the tests. The compare method needs to be updated to handle byte arrays properly.
\end{lstlisting}

%% file: motivating_examples/example_1/15_codegen_attempt_3.tex
\begin{lstlisting}[language=Java]
import sleeper.core.record.Record;
import sleeper.core.schema.Schema;
import java.util.Comparator;
import java.util.List;
import java.util.Arrays;

public class RecordComparator implements Comparator {
    private final List rowKeyFieldNames;
    private final List sortKeyFieldNames;
    public RecordComparator(Schema schema) {
        this.rowKeyFieldNames = schema.getRowKeyFieldNames();
        this.sortKeyFieldNames = schema.getSortKeyFieldNames();
    }

    @Override
    public int compare(Record record1, Record record2) {
        for (String key : rowKeyFieldNames) {
            Object value1 = record1.get(key);
            Object value2 = record2.get(key);
            int comparison = compareValues(value1, value2);
            if (comparison != 0) {
                return comparison;
            }
        }
        for (String key : sortKeyFieldNames) {
            Object value1 = record1.get(key);
            Object value2 = record2.get(key);
            int comparison = compareValues(value1, value2);
            if (comparison != 0) {
                return comparison;
            }
        }
        return 0;
    }
    private int compareValues(Object value1, Object value2) {
        if (value1 instanceof byte[] && value2 instanceof byte[]) {
            return Arrays.compare((byte[]) value1, (byte[]) value2);
        } else if (value1 instanceof Comparable && value2 instanceof Comparable) {
            return ((Comparable) value1).compareTo(value2);
        } else {
            throw new IllegalArgumentException(\"Values are not comparable\");
        }
    }
}
\end{lstlisting}

%% file: sections/appendix/prompt_info.tex
\input{sections/appendix/prompt_templates/initial}

\input{sections/appendix/prompt_templates/toolcall}
\input{sections/appendix/prompt_templates/reflection}
\input{sections/appendix/prompt_templates/codegen}

\input{sections/appendix/diffs/example1}
\input{sections/appendix/diffs/example2}

%% file: sections/appendix/prompt_templates/initial.tex
\begin{table}[!htp]
\centering
\small 

\begin{tabular}{p{\linewidth}}
\begin{lstlisting}[frame=trbl, backgroundcolor=\color{light-gray}, rulecolor=\color{light-gray}]
Solve the below class-generation tasks (include all necessary imports):
# Question 1
Below are some referential code fragments from other files.
{FS_EXAMPLE_SNIPPETS}
Based on the above, generate the following class
File: {FS_EXAMPLE_FILE_PATH}
Description: {FS_EXAMPLE_DESCRIPTION}
Generated Code:
```{LANGUAGE}
{FS_EXAMPLE_CODE}
```
# Question 2
Below are some referential code fragments from other files.
{SNIPPETS}
Based on the above, generate the following class
File: {FILE_PATH}
Description: {DESCRIPTION}
Generated Code:
```{LANGUAGE}
\end{lstlisting}
\end{tabular}
\caption{This is the first prompt that is used fed to the LLM. The intent here is code generation, using a provided description and a few relevant code snippets retrieved from the repository. Relevance is measured as cosine similarity between computed UniXCoder \cite{UniXCoder} embeddings.}
\label{prompts:initial}
\end{table}

%% file: sections/appendix/prompt_templates/toolcall.tex
\begin{table}[!htp]
\centering
\small 
\begin{tabular}{p{\linewidth}}
\begin{lstlisting}[backgroundcolor=\color{light-gray}, rulecolor=\color{light-gray}]
You are a {LANGUAGE} coding assistant. Fix the error in the code by interleaving Thought and Action. `Thought` can be used to reason about the current situation/error. You have been also provided a set of tools/actions to get information about the various parts of the repository. Here is a list of available actions/tools:
(1) get_class_info(class_name): retrieves a list of available methods or properties for a given class `class_name` if it exists. Also returns the constructor.
(2) get_signature(class_name, method_name): which returns the signature of the specified method `method_name` in the class `class_name`, including its parameter names and types, if it exists. 
(3) get_method_body(class_name, method_name): which returns the body of the specified method if it exists. 
(4) get_relevant_code(search_string): which returns potentially relevant pieces of code from the repository corresponding to the `search_string`.
(5) get_imports(): which returns import suggestions for all the undefined symbols in the code.

These are the only permitted actions/tools. Note that you cannot call these actions on the class that needs to be generated since it does not exist yet. That the same tool can also be called multiple times with different arguments. 

Here are some examples.

# Question 1:
File: {FS_EXAMPLE_FILE_PATH}
Description: {FS_EXAMPLE_DESCRIPTION}"
Previous faulty implementation:
```{LANGUAGE}
{FS_EXAMPLE_PREVIOUS_IMPL}
```
Feedback from previous faulty implementation:
{FS_EXAMPLE_PREVIOUS_IMPL_FEEDBACK}

Available tools (the same tool can be called more than once but atmax 3 tool calls can be made): get_class_info(class_name) get_signature(class_name, method_name) get_method_body(class_name, method_name) get_relevant_code(search_string) get_imports()
{fs_example_previous_impl_tool_call}

# Question 2:
File: {FILE_PATH}
Description: {DESCRIPTION}
Previous faulty implementation:
```{LANGUAGE}
{PREVIOUS_IMPL}
```
Feedback from previous faulty implementation:
{PREVIOUS_IMPL_FEEDBACK}

Available tools (the same tool can be called more than once but at max 3 tool calls can be made): get_class_info(class_name), get_signature(class_name, method_name), get_method_body(class_name, method_name), get_relevant_code(search_string) get_imports()
<tool_usage_thoughts>    
\end{lstlisting}

\end{tabular}
\caption{Prompt used to ask the LLM what tools it would like to invoke based on the prior code it generated and the oracle feedback it received for it's generation. Additionally, language specific, curated few-shot examples are provided to act as a template of the expected structure of the output. }
\label{prompts:toolcall}
\end{table}

%% file: sections/appendix/prompt_templates/reflection.tex
\begin{table}[!htp]
\centering
\small 
\begin{tabular}{p{\linewidth}}
\begin{lstlisting}[frame=trbl, backgroundcolor=\color{light-gray}, framesep=10pt, rulecolor=\color{light-gray}, xleftmargin=20pt, xrightmargin=20pt]
You are a {LANGUAGE} writing assistant. You are supposed to generate a {LANGUAGE} class based on the description of the class provided to you. You will be given your previous faulty implementation and feedback about what went wrong based on a few test cases. Your goal is to write a few sentences to explain why your implementation is wrong as indicated by the tests. You will need this as a hint when you try again later. Only provide the few sentence description in your answer, not the implementation. 

# Question 1:
Below are some potentially relevant pieces of information.
{FS_EXAMPLE_TOOL_OBSERVATIONS}
Based on the above, generate the following class
File: {FS_EXAMPLE_FILE_PATH}
Description: {FS_EXAMPLE_DESCRIPTION}
Previous faulty implementation:
```{LANGUAGE}
{FS_EXAMPLE_PREVIOUS_IMPL_V2}
```
Feedback from previous faulty implementation:
{FS_EXAMPLE_PREVIOUS_IMPL_FEEDBACK_V2}
Reflection on previous faulty implementation:
```{FS_EXAMPLE_PREVIOUS_IMPL_REFLECTION_V2}```

# Question 2:
Below are some potentially relevant pieces of information.
{TOOL_OBSERVATIONS}
Based on the above, generate the following class
File: {FILE_PATH}
Description: {DESCRIPTION}
Previous faulty implementation:
```{LANGUAGE}
{PREVIOUS_IMPL}
```
Feedback from previous faulty implementation:
{PREVIOUS_IMPL_FEEDBACK}
Reflection on previous faulty implementation:```
"""    
\end{lstlisting}
\end{tabular}
\caption{Prompt used where the LLM is asked to reflect on how it's generation is faulty based on the oracle feedback.}
\label{prompts:reflection}
\end{table}

%% file: sections/appendix/prompt_templates/codegen.tex
\begin{table}[!htp]
\centering
\small 

\begin{tabular}{p{\linewidth}}
\begin{lstlisting}[frame=trbl, backgroundcolor=\color{light-gray}, rulecolor=\color{light-gray}]
You are a {LANGUAGE} writing assistant. You are supposed to generate a {LANGUAGE} class based on the description of the class provided to you. You will be given your previous faulty implementation, feedback about what went wrong based on a few test cases, and a hint to change the implementation appropriately. Use these to provide a correct implementation of the class (include all necessary imports).

# Question 1:
Below are some potentially relevant pieces of information.
{FS_EXAMPLE_TOOL_OBSERVATIONS}
Based on the above, generate the following class
File: {FS_EXAMPLE_FILE_PATH}
Description: {FS_EXAMPLE_DESCRIPTION}
Previous faulty implementation:
```{LANGUAGE}
{FS_EXAMPLE_PREVIOUS_IMPL_V2}
```
Feedback from previous faulty implementation:
{FS_EXAMPLE_PREVIOUS_IMPL_FEEDBACK_V2}
Reflection on previous faulty implementation:
{FS_EXAMPLE_PREVIOUS_IMPL_REFLECTION_V2}
Corrected code based on feedback and reflection:
```{LANGUAGE}
{FS_EXAMPLE_CODE}
```
# Question 2:
Below are some potentially relevant pieces of information.
{TOOL_OBSERVATIONS}
Based on the above, generate the following class
File: {FILE_PATH}
Description: {DESCRIPTION}
Previous faulty implementation:
```{LANGUAGE}
{PREVIOUS_IMPL}
```
Feedback from previous faulty implementation:
{PREVIOUS_IMPL_FEEDBACK}
Reflection on previous faulty implementation:
{PREVIOUS_IMPL_REFLECTION}
Corrected code based on feedback and reflection:
```{LANGUAGE}"""
\end{lstlisting}

\end{tabular}
\caption{Prompt used for code-generation}
\label{prompts:codegen}
\end{table}

%% file: sections/appendix/diffs/example1.tex
 \begin{table}[!htp]
\centering
\small 
\begin{tabular}{p{\linewidth}}
% \hline
\begin{lstlisting}[language=diff, 
breaklines=true,basicstyle=\tiny\ttfamily,label={diffs:example1}]
--- file1_before.txt    2024-03-30 03:21:05.716952669 -0700
+++ file1_after.txt     2024-03-30 03:11:00.743386767 -0700
@@ -1,66 +1,92 @@
 Class signature: class Accept:
 Class full name: litestar.datastructures.headers.Accept
 
 Functions accessible:
 <Function details for function no. 0>
 Function signature: def __init__(self, accept_value: str) -> None:
 Function fqdn: litestar.datastructures.headers.Accept.__init__
 Decorators:
 Function docstring: Initialize the Accept header with an accept_value.
+Function body: 
+def __init__(self, accept_value: str):
+   self._accepted_types =\
+    [MTHeader(t) for t\
+       in accept_value.split(",")]
+   self._accepted_types.sort(key=lambda t:\
+         t.priority)
 </function details>
 
 <Function details for function no. 1>
 Function signature: def best_match(self, provided_types: List[str], default: Optional[str] = None) -> Optional[str]:
 Function fqdn: litestar.datastructures.headers.Accept.best_match
 Decorators:
 Function docstring: """Find the best matching media type for the request.

     Args:
         provided_types: A list of media types that can be provided as a response. These types
                         can contain a wildcard ``*`` character in the main- or subtype part.
         default: The media type that is returned if none of the provided types match.

     Returns:
         The best matching media type. If the matching provided type contains wildcard characters,
             they are replaced with the corresponding part of the accepted type. Otherwise the
             provided type is returned as-is.
 """
+Function body: 
def best_match(self, provided_types: List[str], default: Optional[str] = None) -> Optional[str]:
+   types = [MTHeader(t) for t in provided_types]
+
+    for accepted in self._accepted_types:
+        for provided in types:
+            if provided.match(accepted):
+                # by concrete parts from the provided type
+                result = copy(provided)
+                if result.subtype == "*":
+                    result.subtype = accepted.subtype
+                if result.maintype == "*":
+                    result.maintype = accepted.maintype
+                return str(result)
+    return default
 </function details>
 
 <Function details for function no. 2>
 Function signature: def accepts(self, media_type: str) -> bool:
 Function fqdn: litestar.datastructures.headers.Accept.accepts
 Decorators:
 Function docstring: """Check if the request accepts the specified media type.

     If multiple media types can be provided, it is better to use :func:`best_match`.

     Args:
         media_type: The media type to check for.

     Returns:
         True if the request accepts ``media_type``.
 """
+Function body: 
def accepts(self, media_type: str):
+   return self.best_match([media_type]) \
+     == media_type
 </function details>
 
 <Function details for function no. 3>
 Function signature: def __len__(self) -> int:
 Function fqdn: litestar.datastructures.headers.Accept.__len__
 Decorators:
 Function docstring: Returns the number of accepted types.
+Function body: def __len__(self) -> int:
+    return len(self._accepted_types)
 </function details>
 
 <Function details for function no. 4>
 Function signature: def __getitem__(self, key: int) -> str:
 Function fqdn: litestar.datastructures.headers.Accept.__getitem__
 Decorators:
 Function docstring: Returns the accepted type at the given index.
+Function body: def __getitem__(self, key: int):
+    return str(self._accepted_types[key])
 </function details>

 \end{lstlisting}

\end{tabular}
\caption{Here, the diff represents the difference in the context provided to GPT-4 while generating NL description for the class in the Sketchy and Detailed settings. The green portion represents the extra details added for the prompt for the Detailed setting as compared to the Sketchy setting.}

\end{table}

%% file: sections/appendix/diffs/example2.tex
\begin{table}[!htp]
\centering
\small 
\begin{tabular}{p{\linewidth}}
\begin{lstlisting}[language=diff, label={diffs:example2}]
--- file2_before.py	2024-03-30 03:51:38.882430546 -0700
+++ file2_after.py	2024-03-30 03:51:03.053152249 -0700
@@ -1,2 +1,2 @@
-class SESsiON(SessiONREDIReCTmiXIn):
+class Session(SessionRedirectMixin):
     """A Requests session.
         #: :class:`Session <Session>`.
-        self.headers = DeFauLt_hEAdERS()
+        self.headers = default_headers()
 
@@ -41,3 +50,3 @@
         #: Event-handling hooks.
-        self.hooks = DEFAult_hOOkS()
+        self.hooks = default_hooks()
 
@@ -79,3 +88,3 @@
         #: may be any other ``cookielib.CookieJar`` compatible object.
-        self.cookies = coOKiEJAr_fROM_dIct({})
+        self.cookies = cookiejar_from_dict({})
 
@@ -83,4 +92,4 @@
         self.adapters = OrderedDict()
-        self.mOUNt('https://', htTPAdAPTEr())
-        self.mOUNt('http://', htTPAdAPTEr())
+        self.mount("https://", HTTPAdapter())
+        self.mount("http://", HTTPAdapter())
 
@@ -106,7 +115,7 @@
         if not isinstance(cookies, cookielib.CookieJar):
-            cookies = coOKiEJAr_fROM_dIct(cookies)
+            cookies = cookiejar_from_dict(cookies)
 
         # Merge with session cookies
-        merged_cookies = mERgE_cOOkIeS(
-            mERgE_cOOkIeS(ReqUesTScoOKIejaR(), self.cookies), cookies)
+        merged_cookies = merge_cookies(
+            merge_cookies(RequestsCookieJar(), self.cookies), cookies)
 
@@ -198,3 +209,3 @@
 
-        settings = self.MERGe_EnvironmeNT_SETTinGS(
+        settings = self.merge_environment_settings(
             prep.url, proxies, stream, verify, cert

@@ -221,5 +232,5 @@
         kwargs.setdefault('allow_redirects', True)
-        return self.REQUest('GET', url, **kwargs)
+        return self.request('GET', url, **kwargs)
 
-    def oPtIOnS(self, url, **kwargs):
+    def options(self, url, **kwargs):
         r"""Sends a OPTIONS request. Returns :class:`Response` object.
 
@@ -373,17 +384,18 @@
             no_proxy = proxies.get('no_proxy') if proxies is not None else None
-            env_proxies = Get_ENviRON_pROxieS(url, no_proxy=no_proxy)
+            env_proxies = get_environ_proxies(url, no_proxy=no_proxy)
                 proxies.setdefault(k, v)
 
         # Merge all the kwargs.
-        proxies = mErGE_sEtTiNG(proxies, self.proxies)
-        stream = mErGE_sEtTiNG(stream, self.stream)
-        verify = mErGE_sEtTiNG(verify, self.verify)
+        proxies = merge_setting(proxies, self.proxies)
+        stream = merge_setting(stream, self.stream)
+        verify = merge_setting(verify, self.verify)
 
\end{lstlisting}
\end{tabular}
\caption{A diff file showing the changes in the body of a candidate class before a symbol-renaming based paraphrase attempt.}

\end{table}

%% file: main.bbl
\begin{thebibliography}{26}
\providecommand{\natexlab}[1]{#1}
\providecommand{\url}[1]{\texttt{#1}}
\expandafter\ifx\csname urlstyle\endcsname\relax
  \providecommand{\doi}[1]{doi: #1}\else
  \providecommand{\doi}{doi: \begingroup \urlstyle{rm}\Url}\fi

\bibitem[Agrawal et~al.(2023)Agrawal, Kanade, Goyal, Lahiri, and Rajamani]{bench:mgd}
Agrawal, L., Kanade, A., Goyal, N., Lahiri, S.~K., and Rajamani, S.
\newblock Monitor-guided decoding of code {LM}s with static analysis of repository context.
\newblock In \emph{Thirty-seventh Conference on Neural Information Processing Systems}, 2023.
\newblock URL \url{https://openreview.net/forum?id=qPUbKxKvXq}.

\bibitem[Athiwaratkun et~al.(2022)Athiwaratkun, Gouda, Wang, Li, Tian, Tan, Ahmad, Wang, Sun, Shang, Gonugondla, Ding, Kumar, Fulton, Farahani, Jain, Giaquinto, Qian, Ramanathan, Nallapati, Ray, Bhatia, Sengupta, Roth, and Xiang]{mxeval}
Athiwaratkun, B., Gouda, S.~K., Wang, Z., Li, X., Tian, Y., Tan, M., Ahmad, W.~U., Wang, S., Sun, Q., Shang, M., Gonugondla, S.~K., Ding, H., Kumar, V., Fulton, N., Farahani, A., Jain, S., Giaquinto, R., Qian, H., Ramanathan, M.~K., Nallapati, R., Ray, B., Bhatia, P., Sengupta, S., Roth, D., and Xiang, B.
\newblock Multi-lingual evaluation of code generation models.
\newblock 2022.
\newblock \doi{10.48550/ARXIV.2210.14868}.
\newblock URL \url{https://arxiv.org/abs/2210.14868}.

\bibitem[Chen et~al.(2021)Chen, Tworek, Jun, Yuan, de~Oliveira~Pinto, Kaplan, Edwards, Burda, Joseph, Brockman, Ray, Puri, Krueger, Petrov, Khlaaf, Sastry, Mishkin, Chan, Gray, Ryder, Pavlov, Power, Kaiser, Bavarian, Winter, Tillet, Such, Cummings, Plappert, Chantzis, Barnes, Herbert-Voss, Guss, Nichol, Paino, Tezak, Tang, Babuschkin, Balaji, Jain, Saunders, Hesse, Carr, Leike, Achiam, Misra, Morikawa, Radford, Knight, Brundage, Murati, Mayer, Welinder, McGrew, Amodei, McCandlish, Sutskever, and Zaremba]{bench:humaneval}
Chen, M., Tworek, J., Jun, H., Yuan, Q., de~Oliveira~Pinto, H.~P., Kaplan, J., Edwards, H., Burda, Y., Joseph, N., Brockman, G., Ray, A., Puri, R., Krueger, G., Petrov, M., Khlaaf, H., Sastry, G., Mishkin, P., Chan, B., Gray, S., Ryder, N., Pavlov, M., Power, A., Kaiser, L., Bavarian, M., Winter, C., Tillet, P., Such, F.~P., Cummings, D., Plappert, M., Chantzis, F., Barnes, E., Herbert-Voss, A., Guss, W.~H., Nichol, A., Paino, A., Tezak, N., Tang, J., Babuschkin, I., Balaji, S., Jain, S., Saunders, W., Hesse, C., Carr, A.~N., Leike, J., Achiam, J., Misra, V., Morikawa, E., Radford, A., Knight, M., Brundage, M., Murati, M., Mayer, K., Welinder, P., McGrew, B., Amodei, D., McCandlish, S., Sutskever, I., and Zaremba, W.
\newblock Evaluating large language models trained on code.
\newblock 2021.

\bibitem[Du et~al.(2023)Du, Liu, Wang, Wang, Liu, Chen, Feng, Sha, Peng, and Lou]{bench:classeval}
Du, X., Liu, M., Wang, K., Wang, H., Liu, J., Chen, Y., Feng, J., Sha, C., Peng, X., and Lou, Y.
\newblock Classeval: A manually-crafted benchmark for evaluating llms on class-level code generation, 2023.

\bibitem[Guo et~al.(2022)Guo, Lu, Duan, Wang, Zhou, and Yin]{UniXCoder}
Guo, D., Lu, S., Duan, N., Wang, Y., Zhou, M., and Yin, J.
\newblock Unixcoder: Unified cross-modal pre-training for code representation, 2022.

\bibitem[Guo et~al.(2024)Guo, Zhu, Yang, Xie, Dong, Zhang, Chen, Bi, Wu, Li, Luo, Xiong, and Liang]{llms:deepseekcoder}
Guo, D., Zhu, Q., Yang, D., Xie, Z., Dong, K., Zhang, W., Chen, G., Bi, X., Wu, Y., Li, Y., Luo, F., Xiong, Y., and Liang, W.
\newblock Deepseek-coder: When the large language model meets programming -- the rise of code intelligence, 2024.
\newblock URL \url{https://arxiv.org/abs/2401.14196}.

\bibitem[Guu et~al.(2020)Guu, Lee, Tung, Pasupat, and Chang]{cr:realm}
Guu, K., Lee, K., Tung, Z., Pasupat, P., and Chang, M.-W.
\newblock Realm: retrieval-augmented language model pre-training.
\newblock In \emph{Proceedings of the 37th International Conference on Machine Learning}, ICML'20. JMLR.org, 2020.

\bibitem[Izacard et~al.(2022)Izacard, Lewis, Lomeli, Hosseini, Petroni, Schick, Dwivedi-Yu, Joulin, Riedel, and Grave]{cr:atlas}
Izacard, G., Lewis, P., Lomeli, M., Hosseini, L., Petroni, F., Schick, T., Dwivedi-Yu, J., Joulin, A., Riedel, S., and Grave, E.
\newblock Few-shot {Learning} with {Retrieval} {Augmented} {Language} {Models}.
\newblock 2022.
\newblock URL \url{http://arxiv.org/abs/2208.03299}.

\bibitem[Jiang et~al.(2023)Jiang, Xu, Gao, Sun, Liu, Dwivedi-Yu, Yang, Callan, and Neubig]{cr:flare}
Jiang, Z., Xu, F.~F., Gao, L., Sun, Z., Liu, Q., Dwivedi-Yu, J., Yang, Y., Callan, J., and Neubig, G.
\newblock Active retrieval augmented generation.
\newblock 2023.

\bibitem[Jimenez et~al.(2024)Jimenez, Yang, Wettig, Yao, Pei, Press, and Narasimhan]{swebench}
Jimenez, C.~E., Yang, J., Wettig, A., Yao, S., Pei, K., Press, O., and Narasimhan, K.~R.
\newblock {SWE}-bench: Can language models resolve real-world github issues?
\newblock In \emph{The Twelfth International Conference on Learning Representations}, 2024.
\newblock URL \url{https://openreview.net/forum?id=VTF8yNQM66}.

\bibitem[Li et~al.(2023)Li, Allal, Zi, Muennighoff, Kocetkov, Mou, Marone, Akiki, Li, Chim, Liu, Zheltonozhskii, Zhuo, Wang, Dehaene, Davaadorj, Lamy-Poirier, Monteiro, Shliazhko, Gontier, Meade, Zebaze, Yee, Umapathi, Zhu, Lipkin, Oblokulov, Wang, Murthy, Stillerman, Patel, Abulkhanov, Zocca, Dey, Zhang, Fahmy, Bhattacharyya, Yu, Singh, Luccioni, Villegas, Kunakov, Zhdanov, Romero, Lee, Timor, Ding, Schlesinger, Schoelkopf, Ebert, Dao, Mishra, Gu, Robinson, Anderson, Dolan-Gavitt, Contractor, Reddy, Fried, Bahdanau, Jernite, Ferrandis, Hughes, Wolf, Guha, von Werra, and de~Vries]{llms:starcoder}
Li, R., Allal, L.~B., Zi, Y., Muennighoff, N., Kocetkov, D., Mou, C., Marone, M., Akiki, C., Li, J., Chim, J., Liu, Q., Zheltonozhskii, E., Zhuo, T.~Y., Wang, T., Dehaene, O., Davaadorj, M., Lamy-Poirier, J., Monteiro, J., Shliazhko, O., Gontier, N., Meade, N., Zebaze, A., Yee, M.-H., Umapathi, L.~K., Zhu, J., Lipkin, B., Oblokulov, M., Wang, Z., Murthy, R., Stillerman, J., Patel, S.~S., Abulkhanov, D., Zocca, M., Dey, M., Zhang, Z., Fahmy, N., Bhattacharyya, U., Yu, W., Singh, S., Luccioni, S., Villegas, P., Kunakov, M., Zhdanov, F., Romero, M., Lee, T., Timor, N., Ding, J., Schlesinger, C., Schoelkopf, H., Ebert, J., Dao, T., Mishra, M., Gu, A., Robinson, J., Anderson, C.~J., Dolan-Gavitt, B., Contractor, D., Reddy, S., Fried, D., Bahdanau, D., Jernite, Y., Ferrandis, C.~M., Hughes, S., Wolf, T., Guha, A., von Werra, L., and de~Vries, H.
\newblock Starcoder: may the source be with you!, 2023.

\bibitem[Liu et~al.(2023)Liu, Xu, and McAuley]{bench:repobench}
Liu, T., Xu, C., and McAuley, J.
\newblock Repobench: Benchmarking repository-level code auto-completion systems, 2023.

\bibitem[Luo et~al.(2023)Luo, Xu, Zhao, Sun, Geng, Hu, Tao, Ma, Lin, and Jiang]{llms:wizardcoder}
Luo, Z., Xu, C., Zhao, P., Sun, Q., Geng, X., Hu, W., Tao, C., Ma, J., Lin, Q., and Jiang, D.
\newblock Wizardcoder: Empowering code large language models with evol-instruct, 2023.

\bibitem[Madaan et~al.(2023)Madaan, Tandon, Gupta, Hallinan, Gao, Wiegreffe, Alon, Dziri, Prabhumoye, Yang, Welleck, Majumder, Gupta, Yazdanbakhsh, and Clark]{reasoning:self_refine}
Madaan, A., Tandon, N., Gupta, P., Hallinan, S., Gao, L., Wiegreffe, S., Alon, U., Dziri, N., Prabhumoye, S., Yang, Y., Welleck, S., Majumder, B.~P., Gupta, S., Yazdanbakhsh, A., and Clark, P.
\newblock Self-refine: Iterative refinement with self-feedback, 2023.

\bibitem[Nijkamp et~al.(2023)Nijkamp, Pang, Hayashi, Tu, Wang, Zhou, Savarese, and Xiong]{llms:codegen}
Nijkamp, E., Pang, B., Hayashi, H., Tu, L., Wang, H., Zhou, Y., Savarese, S., and Xiong, C.
\newblock Codegen: An open large language model for code with multi-turn program synthesis.
\newblock \emph{ICLR}, 2023.

\bibitem[Odena et~al.(2021)Odena, Sutton, Dohan, Jiang, Michalewski, Austin, Bosma, Nye, Terry, and Le]{bench:mbpp}
Odena, A., Sutton, C., Dohan, D.~M., Jiang, E., Michalewski, H., Austin, J., Bosma, M.~P., Nye, M., Terry, M., and Le, Q.~V.
\newblock Program synthesis with large language models.
\newblock In \emph{n/a}, pp.\  n/a, n/a, 2021.
\newblock n/a.

\bibitem[Rozière et~al.(2024)Rozière, Gehring, Gloeckle, Sootla, Gat, Tan, Adi, Liu, Sauvestre, Remez, Rapin, Kozhevnikov, Evtimov, Bitton, Bhatt, Ferrer, Grattafiori, Xiong, Défossez, Copet, Azhar, Touvron, Martin, Usunier, Scialom, and Synnaeve]{llms:codellama}
Rozière, B., Gehring, J., Gloeckle, F., Sootla, S., Gat, I., Tan, X.~E., Adi, Y., Liu, J., Sauvestre, R., Remez, T., Rapin, J., Kozhevnikov, A., Evtimov, I., Bitton, J., Bhatt, M., Ferrer, C.~C., Grattafiori, A., Xiong, W., Défossez, A., Copet, J., Azhar, F., Touvron, H., Martin, L., Usunier, N., Scialom, T., and Synnaeve, G.
\newblock Code llama: Open foundation models for code, 2024.

\bibitem[Schick et~al.(2023)Schick, Dwivedi-Yu, Dessì, Raileanu, Lomeli, Zettlemoyer, Cancedda, and Scialom]{tools:toolformer}
Schick, T., Dwivedi-Yu, J., Dessì, R., Raileanu, R., Lomeli, M., Zettlemoyer, L., Cancedda, N., and Scialom, T.
\newblock Toolformer: Language models can teach themselves to use tools, 2023.

\bibitem[Shinn et~al.(2023)Shinn, Cassano, Berman, Gopinath, Narasimhan, and Yao]{reasoning:reflexion}
Shinn, N., Cassano, F., Berman, E., Gopinath, A., Narasimhan, K., and Yao, S.
\newblock Reflexion: Language agents with verbal reinforcement learning, 2023.

\bibitem[Shrivastava et~al.(2023)Shrivastava, Larochelle, and Tarlow]{cr:rlpg}
Shrivastava, D., Larochelle, H., and Tarlow, D.
\newblock Repository-level prompt generation for large language models of code, 2023.

\bibitem[Yao et~al.(2023)Yao, Zhao, Yu, Du, Shafran, Narasimhan, and Cao]{tools:react}
Yao, S., Zhao, J., Yu, D., Du, N., Shafran, I., Narasimhan, K.~R., and Cao, Y.
\newblock React: Synergizing reasoning and acting in language models.
\newblock In \emph{The Eleventh International Conference on Learning Representations}, 2023.
\newblock URL \url{https://openreview.net/forum?id=WE_vluYUL-X}.

\bibitem[Yin et~al.(2018)Yin, Deng, Chen, Vasilescu, and Neubig]{bench:CoNaLA}
Yin, P., Deng, B., Chen, E., Vasilescu, B., and Neubig, G.
\newblock Learning to mine aligned code and natural language pairs from stack overflow.
\newblock In \emph{International Conference on Mining Software Repositories}, MSR, pp.\  476--486. ACM, 2018.
\newblock \doi{https://doi.org/10.1145/3196398.3196408}.

\bibitem[Zhang et~al.(2023)Zhang, Chen, Zhang, Keung, Liu, Zan, Mao, Lou, and Chen]{cr:repocoder}
Zhang, F., Chen, B., Zhang, Y., Keung, J., Liu, J., Zan, D., Mao, Y., Lou, J.-G., and Chen, W.
\newblock {R}epo{C}oder: Repository-level code completion through iterative retrieval and generation.
\newblock In Bouamor, H., Pino, J., and Bali, K. (eds.), \emph{Proceedings of the 2023 Conference on Empirical Methods in Natural Language Processing}, pp.\  2471--2484, Singapore, December 2023. Association for Computational Linguistics.
\newblock \doi{10.18653/v1/2023.emnlp-main.151}.
\newblock URL \url{https://aclanthology.org/2023.emnlp-main.151}.

\bibitem[Zhang et~al.(2021)Zhang, Sun, Gao, Fang, Brockett, Galley, Gao, and Dolan]{cr:retgen}
Zhang, Y., Sun, S., Gao, X., Fang, Y., Brockett, C., Galley, M., Gao, J., and Dolan, B.
\newblock Joint retrieval and generation training for grounded text generation.
\newblock \emph{arXiv preprint arXiv:2105.06597}, 2021.

\bibitem[Zhang et~al.(2024)Zhang, Zhang, Ran, Zhu, Dou, Hao, Xie, and Zhang]{bench:coder_eval}
Zhang, Y., Zhang, W., Ran, D., Zhu, Q., Dou, C., Hao, D., Xie, T., and Zhang, L.
\newblock Learning-based widget matching for migrating gui test cases.
\newblock In \emph{Proceedings of the 46th IEEE/ACM International Conference on Software Engineering}, ICSE ’24. ACM, February 2024.
\newblock \doi{10.1145/3597503.3623322}.
\newblock URL \url{http://dx.doi.org/10.1145/3597503.3623322}.

\bibitem[Zhou et~al.(2023)Zhou, Yan, Shlapentokh-Rothman, Wang, and Wang]{reasoning:lats}
Zhou, A., Yan, K., Shlapentokh-Rothman, M., Wang, H., and Wang, Y.-X.
\newblock Language agent tree search unifies reasoning acting and planning in language models, 2023.

\end{thebibliography}
